\PassOptionsToPackage{unicode}{hyperref}
\PassOptionsToPackage{hyphens}{url}
\documentclass[
]{article}
\usepackage{amsmath,amssymb}
\usepackage{lmodern}
\usepackage{iftex}
\ifPDFTeX
  \usepackage[T1]{fontenc}
  \usepackage[utf8]{inputenc}
  \usepackage{textcomp} 
\else 
  \usepackage{unicode-math}
  \defaultfontfeatures{Scale=MatchLowercase}
  \defaultfontfeatures[\rmfamily]{Ligatures=TeX,Scale=1}
\fi
\IfFileExists{upquote.sty}{\usepackage{upquote}}{}
\IfFileExists{microtype.sty}{
  \usepackage[]{microtype}
  \UseMicrotypeSet[protrusion]{basicmath} 
}{}
\makeatletter
\@ifundefined{KOMAClassName}{
  \IfFileExists{parskip.sty}{%
    \usepackage{parskip}
  }{
    \setlength{\parindent}{0pt}
    \setlength{\parskip}{6pt plus 2pt minus 1pt}}
}{
  \KOMAoptions{parskip=half}}
\makeatother
\usepackage{xcolor}
\usepackage[margin=1in]{geometry}
\usepackage{color}
\usepackage{fancyvrb}

\DefineVerbatimEnvironment{Highlighting}{Verbatim}{commandchars=\\\{\}}
\usepackage{framed}
\definecolor{shadecolor}{RGB}{248,248,248}
\newenvironment{Shaded}{\begin{snugshade}}{\end{snugshade}}

\newcommand{\AttributeTok}[1]{\textcolor[rgb]{0.77,0.63,0.00}{#1}}

\newcommand{\ConstantTok}[1]{\textcolor[rgb]{0.00,0.00,0.00}{#1}}
\newcommand{\ControlFlowTok}[1]{\textcolor[rgb]{0.13,0.29,0.53}{\textbf{#1}}}

\newcommand{\DecValTok}[1]{\textcolor[rgb]{0.00,0.00,0.81}{#1}}
\newcommand{\DocumentationTok}[1]{\textcolor[rgb]{0.56,0.35,0.01}{\textbf{\textit{#1}}}}

\newcommand{\FloatTok}[1]{\textcolor[rgb]{0.00,0.00,0.81}{#1}}
\newcommand{\FunctionTok}[1]{\textcolor[rgb]{0.00,0.00,0.00}{#1}}

\newcommand{\NormalTok}[1]{#1}

\newcommand{\OtherTok}[1]{\textcolor[rgb]{0.56,0.35,0.01}{#1}}

\newcommand{\SpecialCharTok}[1]{\textcolor[rgb]{0.00,0.00,0.00}{#1}}

\newcommand{\StringTok}[1]{\textcolor[rgb]{0.31,0.60,0.02}{#1}}

\usepackage{graphicx}
\makeatletter
\def\maxwidth{\ifdim\Gin@nat@width>\linewidth\linewidth\else\Gin@nat@width\fi}
\def\maxheight{\ifdim\Gin@nat@height>\textheight\textheight\else\Gin@nat@height\fi}
\makeatother
\setkeys{Gin}{width=\maxwidth,height=\maxheight,keepaspectratio}
\makeatletter
\def\fps@figure{htbp}
\makeatother
\providecommand{\tightlist}{%
  \setlength{\itemsep}{0pt}\setlength{\parskip}{0pt}}
\usepackage{amsmath, amsthm, amssymb}
\usepackage{nicefrac}
\usepackage{bm}
\usepackage{setspace}
\usepackage[margin = 1in]{geometry}

\ifLuaTeX
  \usepackage{selnolig}  
\fi
\usepackage[]{natbib}
\IfFileExists{bookmark.sty}{\usepackage{bookmark}}{\usepackage{hyperref}}
\IfFileExists{xurl.sty}{\usepackage{xurl}}{} 
\hypersetup{
  pdftitle={SoftBart: Soft Bayesian Additive Regression Trees},
  pdfauthor={Antonio R. Linero},
  pdfkeywords={true},
  hidelinks,
  pdfcreator={LaTeX via pandoc}}

\title{SoftBart: Soft Bayesian Additive Regression Trees}
\author{Antonio R. Linero\footnote{University of Texas at Austin,
  \href{mailto:antonio.linero@austin.utexas.edu}{\nolinkurl{antonio.linero@austin.utexas.edu}}}}
\date{2022-10-28}

\begin{document}
\maketitle

\begin{abstract}
  Bayesian additive regression tree (BART) models have seen increased attention
  in recent years as a general-purpose nonparametric modeling technique. BART 
  combines the flexibility of modern machine learning techniques with the 
  principled uncertainty quantification of Bayesian inference, and it has been 
  shown to be uniquely appropriate for addressing the high-noise problems that 
  occur commonly in many areas of science, including medicine and the social 
  sciences. This paper introduces the \texttt{SoftBart} package for fitting the 
  \emph{Soft BART} algorithm of \citet{linero2017abayesian}. In addition to 
  improving upon the predictive performance of other BART packages, a major goal 
  of this package has been to facilitate the inclusion of BART in larger models, 
  making it  ideal for researchers in Bayesian statistics. I show both how to
  use this package for standard prediction tasks and how to embed BART models
  in larger models; I illustrate by using \texttt{SoftBart} to implement a
  nonparametric probit regression model, a semiparametric varying coefficient 
  model, and a partial linear model.
\end{abstract}

\hypertarget{introduction}{%
\section{Introduction}\label{introduction}}

\begin{figure}[t]

{\centering \includegraphics[width=0.75\textwidth,]{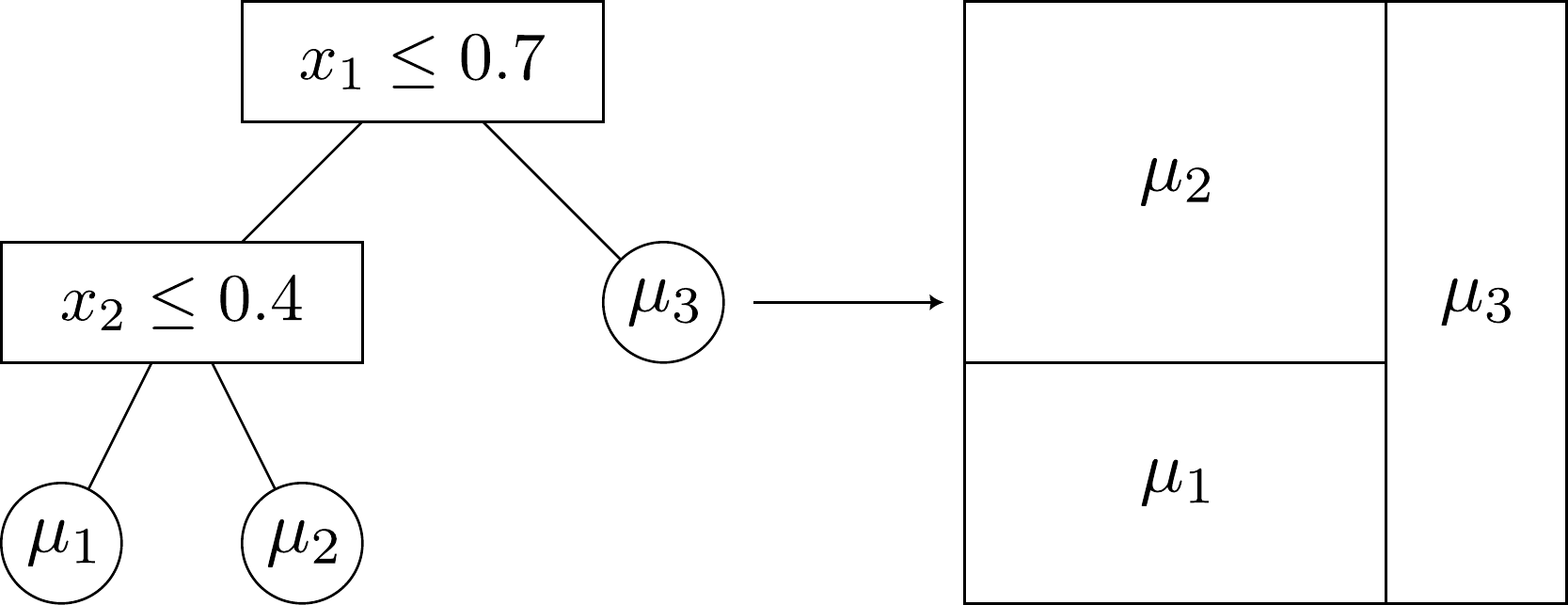} 

}

\caption{Schematic depiction of a decision tree; the left figure gives the decision 
tree $\mathcal T$ and its leaf node values $\mathcal M= (\mu_1, \mu_2, \mu_3)$, while 
the right figure gives the induced step function on $[0,1]^2$. A given value of 
$x$ starts at the top of the tree and goes left if the rule at each node is true
and goes right if the rule is false, until it reaches a terminal node.}\label{fig:treefig}
\end{figure}

Introduced by \citet{chipman2010bart}, Bayesian additive regression tree
(or BART) models have attracted substantial interest from the Bayesian
nonparametrics and machine learning communities. BART is used in
nonparametric function estimation problems, where the function of
interest \(r(x)\) is modeled as a \emph{decision tree ensemble}
(containing \(T\) trees) of the form \[
  r(x) = \sum_{t = 1}^T \operatorname{Tree}(x; \mathcal T_t, \mathcal M_t) 
  \qquad \text{where} \qquad
  (\mathcal T_t, \mathcal M_t) \stackrel{\text{iid}}{\sim}\pi_{\mathcal T}(\mathcal T_t) \, \pi_{\mathcal M}(\mathcal M_t \mid \mathcal T_t),
\] and where \((\pi_\mathcal T, \pi_{\mathcal M})\) defines a prior for
the the parameters of the decision trees. Each of the functions
\(\operatorname{Tree}(x; \mathcal T_t, \mathcal M_t)\) defines a
\emph{regression tree} (see Figure \ref{fig:treefig}). This model can be
viewed as a Bayesian version of the famous \emph{decision tree boosting}
framework \citep{freund1999short, friedman2001greedy}, with the \(T\)
decision trees in the ensemble representing ``weak learners'' that are
then aggregated into a single ``strong learner'' for \(r(x)\). The
canonical problem where BART is applied is the \emph{semiparametric
Gaussian regression} problem \begin{align*}
  Y_i = r(X_i) + \epsilon_i \qquad \text{where} 
      \qquad \epsilon_i \sim \operatorname{Normal}(0, \sigma^2),
\end{align*} but it can also be used to model nonparametric functions
\(r(x)\) in essentially arbitrary problems; these problems include
nonparametric probit and logit regression
\citep{chipman2010bart, murray2021log}, survival analysis
\citep{sparapani2016nonparametric, linero2021bayesian, basak2020semiparametric, henderson2020individualized, bonato2010bayesian},
density regression
\citep{li2020adaptive, orlandi2021density, um2022skew}, and estimation
of inhomogeneous Poisson processes \citep{lamprinakou2020bart}.
Moreover, \citet{linero2022generalized} shows how BART can be used with
arbitrary probabilistic models using reversible jump Markov chain Monte
Carlo, allowing it to be applied in the same settings that one can apply
decision tree boosting.

In this way, BART can be viewed as a flexible drop-in replacement for
other Bayesian nonparametric and/or tree-based function estimation
approaches. In my view, the most important features of the BART
framework are the following:

\begin{enumerate}
\def\labelenumi{\arabic{enumi}.}
\item
  Among Bayesian nonparametric approaches, BART is unique in that there
  are several high-quality, easy to use, software implementations,
  including the \texttt{BayesTree} \citep{chipman2016bayes},
  \texttt{bartMachine} \citep{kapelner2014bartmachine}, \texttt{dbarts}
  \citep{dorie2022dbarts}, and \texttt{BART} \citep{sparapani2021bart}
  packages. These mainly focus on the problems of semiparametric
  Gaussian and nonparametric binary regression. Because of the existence
  of these packages (and the common practice in the BART community of
  proposing default priors) less expertise is required to use BART than
  other Bayesian nonparametric models such as Gaussian processes
  \citep{rasmussen2005gaussian} or Dirichlet process mixtures
  \citep{escobarwest1995}.
\item
  BART has been shown to perform extremely well in the high-noise
  settings that are typical in the social sciences and medicine; for
  example \citet{linero2017abayesian} showed that BART routinely
  outperformed both random forests and decision tree boosting on average
  over many datasets. Because of this, BART has seen wide deployment in
  the causal inference literature
  \citep{hahn2020bayesian, hill2011bayesian, linero2022mediation}, where
  it is used both for Bayesian estimation of heterogeneous causal
  effects and as a black-box machine learning algorithm by Frequentists.
\item
  Unlike most other black-box machine learning methods (e.g., neural
  networks, Gaussian processes, and random forests), the BART prior
  \emph{shrinks towards low-order interactions} in the data --- because
  the decision trees used by BART are usually shallow, realizations of
  \(r(x)\) from the prior will tend to prioritize main effects, with a
  smaller number of second-order interactions, and even fewer high-order
  interactions. While complex high-order interactions are common in some
  fields (image recognition, natural language processing, and so forth),
  they are not thought to be very important in traditional areas of
  statistics. That BART emphasizes low-order interactions was part of
  the initial motivation of \citet{chipman2010bart}, as that BART is
  optimal in this regime was established theoretically in a series of
  recent papers
  \citep{rockova2017posterior, linero2017abayesian, saha2021flexible}.
\end{enumerate}

\begin{figure}[t]

{\centering \includegraphics[width=0.9\textwidth,]{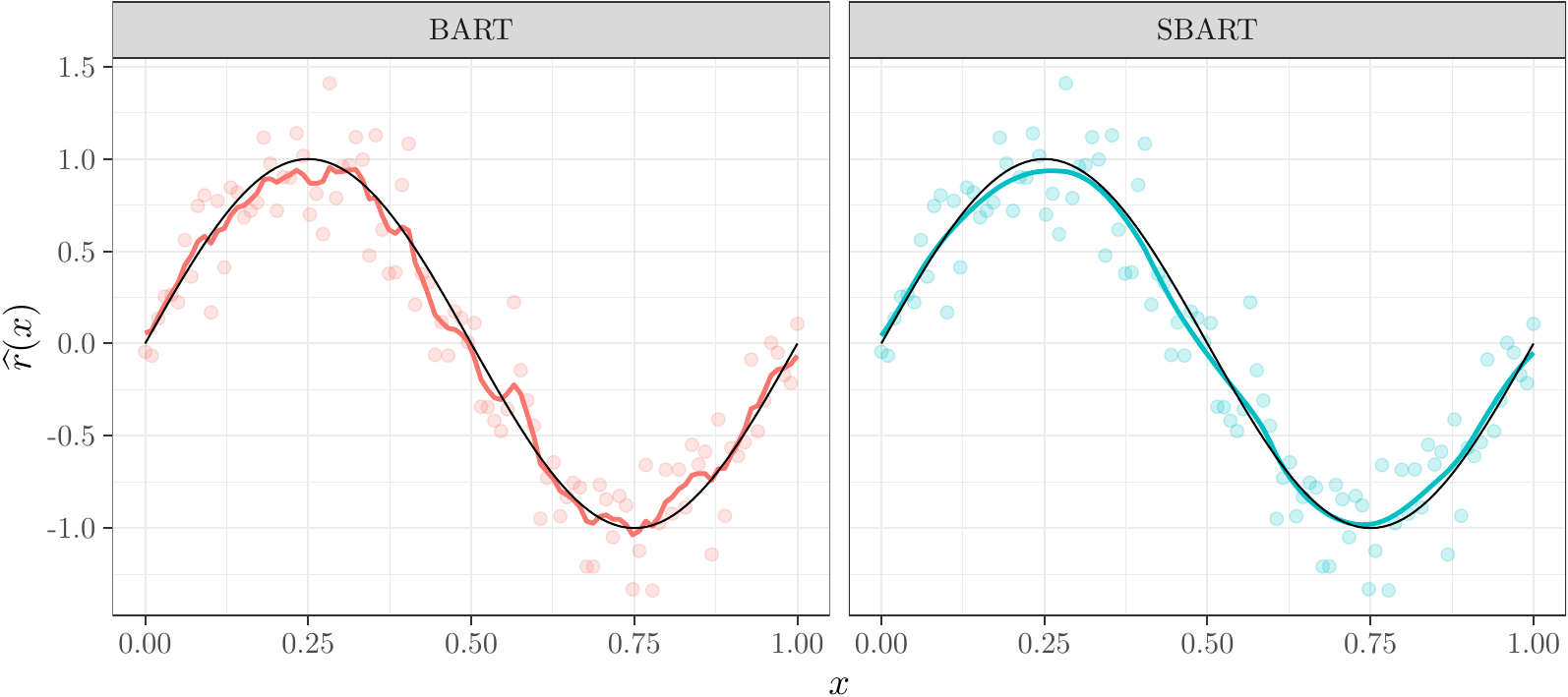} 

}

\caption{Comparison of the fit of the BART and \texttt{SoftBart} models to data generated from the relationship $Y_i = \sin(2 \pi x) + \epsilon_i$ with $\epsilon_i \sim \operatorname{Normal}(0, 0.1^2)$. The sine curve is overlayed in black.}\label{fig:sineplot}
\end{figure}

To balance these positives, BART has several shortcomings. One
shortcoming of BART, and of decision tree ensembling approaches in
general, is that the predictions produced by these models are generally
non-smooth; for example, realizations from the BART prior are
stepwise-continuous functions. The impact of this lack of smoothness can
be seen in Figure \ref{fig:sineplot}, where BART performs suboptimally
in estimating the univariate function \(r(x) = \sin(2\pi x)\). In this
case, the mean-squared error of the BART model is 140\% larger than the
mean-squared error of the \texttt{SoftBart} model we propose here.

To address the lack of smoothness of BART, \citet{linero2017abayesian}
introduced the \texttt{SoftBart} model, with the authors demonstrating
both theoretically (through studies of posterior concentration rates)
and practically (through the analysis of benchmark datasets) that
leveraging smoothness often results in substantially improved prediction
on real datasets. Since its introduction, the \texttt{SoftBart} model
has been used by many researchers; in addition to extensions proposed by
its progenitors, it has seen substantial interest in both the Bayesian
nonparametrics and the Bayesian causal inference literature. A
non-comprehensive list of applications include:
\citet{liu2021inference}, who use \texttt{SoftBart} as the algorithm of
choice for addressing non-response in surveys;
\citet{ran2021distributed}, who constructed a MAP-reduce algorithm for
fitting \texttt{SoftBart} to massive datasets; \citet{bai2022spike}, who
found \texttt{SoftBart} to be a very high-quality competitor to their
spike-and-slab group lasso GAM model; and \citet{hahn2020bayesian},
where the method was discussed both in the main manuscript and by
multiple discussants.

\hypertarget{our-motivation}{%
\subsection{Our Motivation}\label{our-motivation}}

In this paper, we introduce the \texttt{SoftBart} package, and show how
to apply it to a number of nonparametric estimation problems. Given the
large number of high-quality packages for implementing BART, as well as
the large number of competing nonparametric techniques (Bayesian or
otherwise), it is natural to wonder at the value added by yet another
package. There are two major goals of this package:

\begin{enumerate}
\def\labelenumi{\arabic{enumi}.}
\item
  this package makes accessible the \texttt{SoftBart} methodology of
  \citet{linero2017abayesian} to a wider audience; the value of this is
  apparent from the analysis of benchmark datasets given by
  \citet{linero2017abayesian}, as well as the fact that
  \texttt{SoftBart} has been consistently observed to outperform other
  variants of BART by other authors \citep{prado2021bayesian}.
\item
  this package includes functionality that makes it easy to embed BART
  (or \texttt{SoftBart}) into larger \texttt{R} programs; this will aide
  researchers interested in BART by making it easy to include BART in
  custom models.
\end{enumerate}

The second objective is a novel contribution of this package, as to the
best of my knowledge \texttt{SoftBart} is the only BART package that
allows users to embed BART within a larger model without having to
modify the underlying \texttt{C++} code. To this point, work extending
BART has mainly been carried out by, as most statisticians do not have
the requisite programming knowledge (such as familiarity with
\texttt{C++} or other compiled languages, knowledge of tree-based data
structures, and experience implementing Markov chain Monte Carlo
algorithms on discrete data structures) to modify existing BART code
effectively. Part of the motivation for developing this functionality
comes, in fact, from the realities of working on projects with graduate
students; for many problems I have worked on, the modifications required
were conceptually simple, but nevertheless required the student to learn
a non-trivial amount of \texttt{C++} to implement. Making extending BART
more convenient for graduate students was therefore an important aim for
this package.

I show through several illustrations how \texttt{SoftBart} makes it easy
for non-experts to implement extensions of BART; all that is needed is a
conceptual understanding of Gibbs sampling and experience implementing
MCMC algorithms in \texttt{R}. By embedding the BART updates inside of
simple Gibbs sampling algorithms, I show how to implement the following
models: a nonparametric probit regression BART model described by
\citet{chipman2010bart}; the varying coefficient model of
\citet{deshpande2020vcbart}, which I show also contains the Bayesian
causal forests model of \citet{hahn2020bayesian} as a special case; and
the general BART model of \citet{tan2019bayesian}. Beyond these models,
I also note here that several existing works have used a preliminary
version of this software to implement their proposed methodology,
including the survival analysis model of \citet{basak2020semiparametric}
and the skewed error model of \citet{um2022skew}.

\hypertarget{description-of-methodology}{%
\subsection{Description of
Methodology}\label{description-of-methodology}}

We begin by reviewing the original BART model of \citet{chipman2010bart}
before describing the \texttt{SoftBart} model. The BART framework models
an unknown function \(r(x)\) as a \emph{sum of decision trees}
\begin{align}
  \label{eq:sum-of-trees}
  r(x) = \sum_{t=1}^T \operatorname{Tree}(x ; \mathcal T_t, \mathcal M_t)
\end{align} where \(\mathcal T_t\) denotes a \emph{decision tree},
\(\mathcal M_t\) denotes a collection of \emph{leaf node parameters},
and \(\operatorname{Tree}(x ; \mathcal T, \mathcal M)\) is a
\emph{regression tree function} that returns the prediction associated
to \(x\) for the pair \((\mathcal T, \mathcal M)\). I illustrate this in
Figure \ref{fig:treefig}, which gives a schematic of a decision tree
\(\mathcal T\) with leaf node parameters
\(\mathcal M= \{\mu_1, \mu_2, \mu_3\}\); for example, if
\(x = (0.3, 0.6)^\top\) then
\(\operatorname{Tree}(x; \mathcal T, \mathcal M) = \mu_2\) because
\([0.3 \le 0.7]\) and \([0.6 > 0.4]\). Figure \ref{fig:treefig} also
shows that each regression tree
\(\operatorname{Tree}(x; \mathcal T, \mathcal M)\) corresponds to a
piecewise constant function.

As seen in Figure \ref{fig:treefig}, each tree \(\mathcal T\) consists
of a collection of \emph{leaf nodes} (the nodes with no ``child'' nodes)
and a collection of \emph{branch nodes} that each have an associated
\emph{splitting rule} of the form \([x_{j_b} \le C_b]\). We write
\(\mathcal L(\mathcal T)\) for the collection of leaf nodes \(\ell\) and
write \(\mathcal B(\mathcal T)\) for the collection of branch nodes
\(b\).

\hypertarget{the-prior-on-decision-trees}{%
\subsection{The Prior on Decision
Trees}\label{the-prior-on-decision-trees}}

Unlike other decision tree ensembling strategies, such as random forests
or boosting \citep{breiman2001random, freund1999short}, BART proceeds by
placing a prior on the regression trees. Given the model hyperparameters
\(\vartheta = (s, \beta, \gamma, T, \sigma^2_\mu)\) (to be described
later), BART takes the regression trees to be a-priori independent,
i.e., \[
  \pi\big((\mathcal T_1, \mathcal M_1), \ldots, (\mathcal T_T, \mathcal M_T) \mid \vartheta \big)
  =
  \prod_{t = 1}^T \pi_\mathcal T(\mathcal T_t \mid \vartheta) \, \pi_\mathcal M(\mathcal M_t \mid \mathcal T_t).
\] In \texttt{SoftBart}, the prior distribution for the trees
\(\pi_\mathcal T\) consists of two components:

\begin{enumerate}
\def\labelenumi{\arabic{enumi}.}
\tightlist
\item
  a prior on the shape of the tree \(\mathcal T\); and
\item
  a prior on the splitting rules \([x_{j_b} \le C_b]\) for each branch
  node of the tree.
\end{enumerate}

The prior on the shape of \(\mathcal T\) is a \emph{branching process}
described by \citet{chipman2010bart}. We start with a tree consisting
only of a root node of depth \(d = 0\); we then make this root a branch
node with two children with probability \begin{align}
  \label{eq:tree-split-prob}
  \Pr(\text{is a branch}) = \frac{\gamma}{(1 + d)^\beta}
\end{align} otherwise the root becomes a leaf node. This process then
iterates for \(d = 1, 2, \ldots\) until all nodes at a given depth are
leaf nodes.

\texttt{SoftBart} uses a prior for the splitting rules that first
selects a predictor by sampling
\(j_b \sim \operatorname{Categorical}(s)\) where
\(s = (s_1, \ldots, s_P)^\top\) is a probability vector. The prior then
selects the cutpoint \(C_b\) by sampling
\(C_b \sim \operatorname{Uniform}(A_{j_b}, B_{j_b})\) where
\(\prod_{j=1}^P [A_j, B_j]\) is the hyperrectangle of \(x\) values that
are associated to branch \(b\).

The hyperparameters \((s, \gamma, \beta, \sigma_\mu, T)\) are, by
default, either fixed or given weakly-informative hyperpriors, but users
can specify their own values/priors for these quantities if desired. We
defer discussion of the prior on \(s\) to Section
\ref{variable-selection}, as \(s\) plays an important role in variable
selection. \texttt{SoftBart} follows the convention of
\citet{chipman2010bart} by taking \(\gamma = 0.95\) and \(\beta = 2\) by
default, and opts for using fewer trees (\(T = 20\)) by default than
other BART packages. Finally, \texttt{SoftBart} uses the a half-Cauchy
prior \(\sigma_\mu \sim \operatorname{Cauchy}_+(\widehat\sigma_\mu)\)
for \(\sigma_\mu\), where \(\widehat\sigma_\mu = 0.5 / (k \sqrt{T})\)
and \(k = 2\). This is different than other BART packages in that we use
a hyperprior for \(\sigma_\mu\), but the prior is centered at the
default value of \(\sigma_\mu\) recommended for semiparametric Gaussian
regression by \citet{chipman2010bart} (after scaling the outcome \(Y_i\)
to lie in the interval \([-0.5, 0.5]\)). For BART models other than the
semiparametric Gaussian regression model, this choice of \(\sigma_\mu\)
might not be appropriate, and we use (for example) the default
\(\widehat\sigma_\mu = 3 / \sqrt{T}\) for the probit regression model
discussed in Section \ref{probit-regression-with-data-augmentation}.

The prior used in \texttt{SoftBart} differs in several minor ways from
the prior described by \citet{chipman2010bart}. First, we use continuous
uniform cutpoints rather than taking the cutpoints to occur only at the
observed values of the \(X_{ij}\)'s. Second, we do not place any
restrictions on the number of \(X_i\)'s required for a node to be made a
branch; the high level motivation for such restrictions is to reduce the
risk of overfitting, but I have found it to be redundant in practice
given that the prior regularizes the predictions at the leaf nodes.
Ultimately, I have found that these differences make very little
difference in practice in terms of predictive performance, and my choice
of prior is driven by other concerns (in particular, it allows for a
simple conditionally conjugate prior for \(s\)).

\hypertarget{smoothing-decision-trees}{%
\subsection{Smoothing Decision Trees}\label{smoothing-decision-trees}}

\begin{figure}[t]

{\centering \includegraphics[width=0.45\textwidth,]{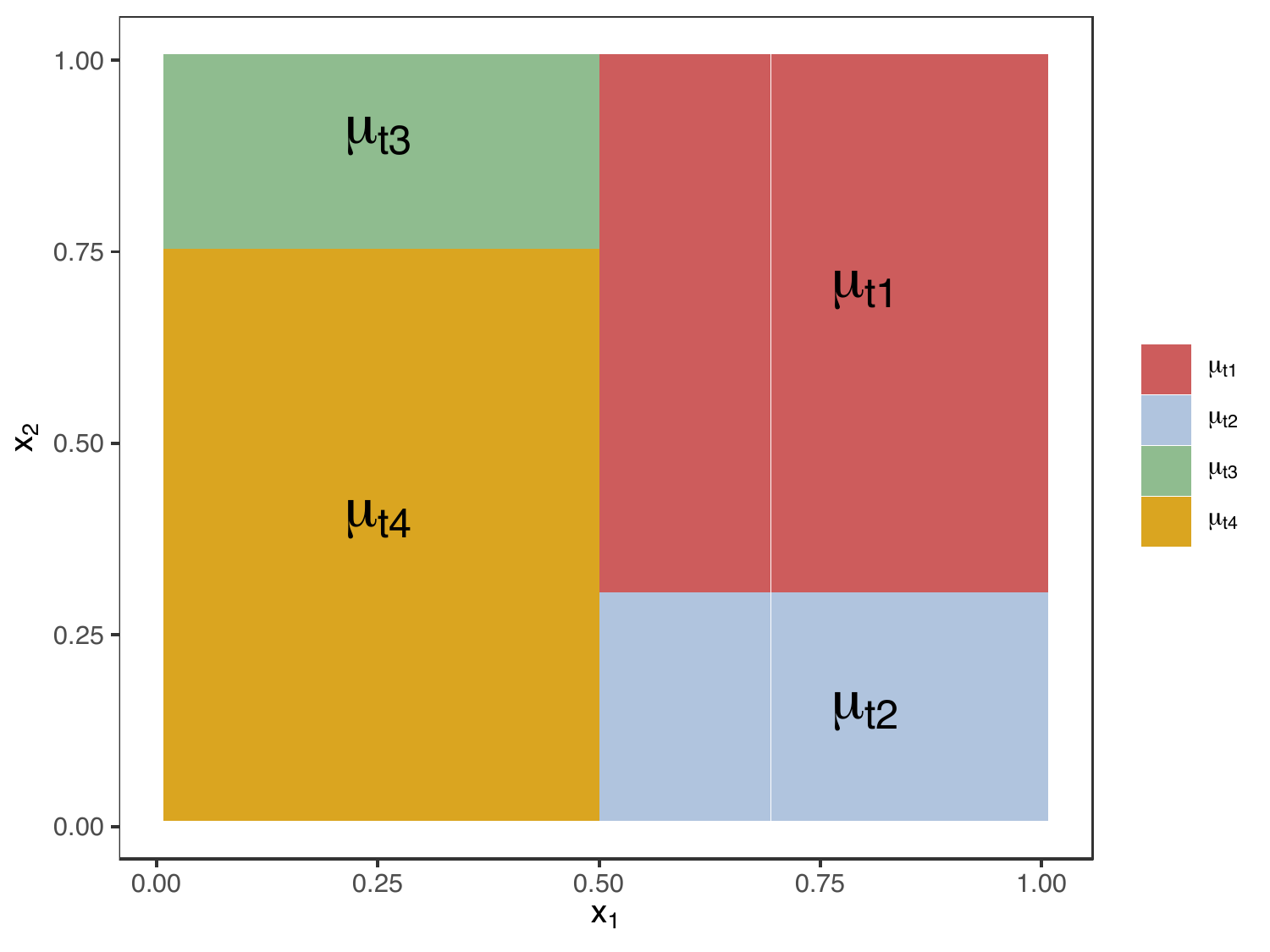} \includegraphics[width=0.45\textwidth,]{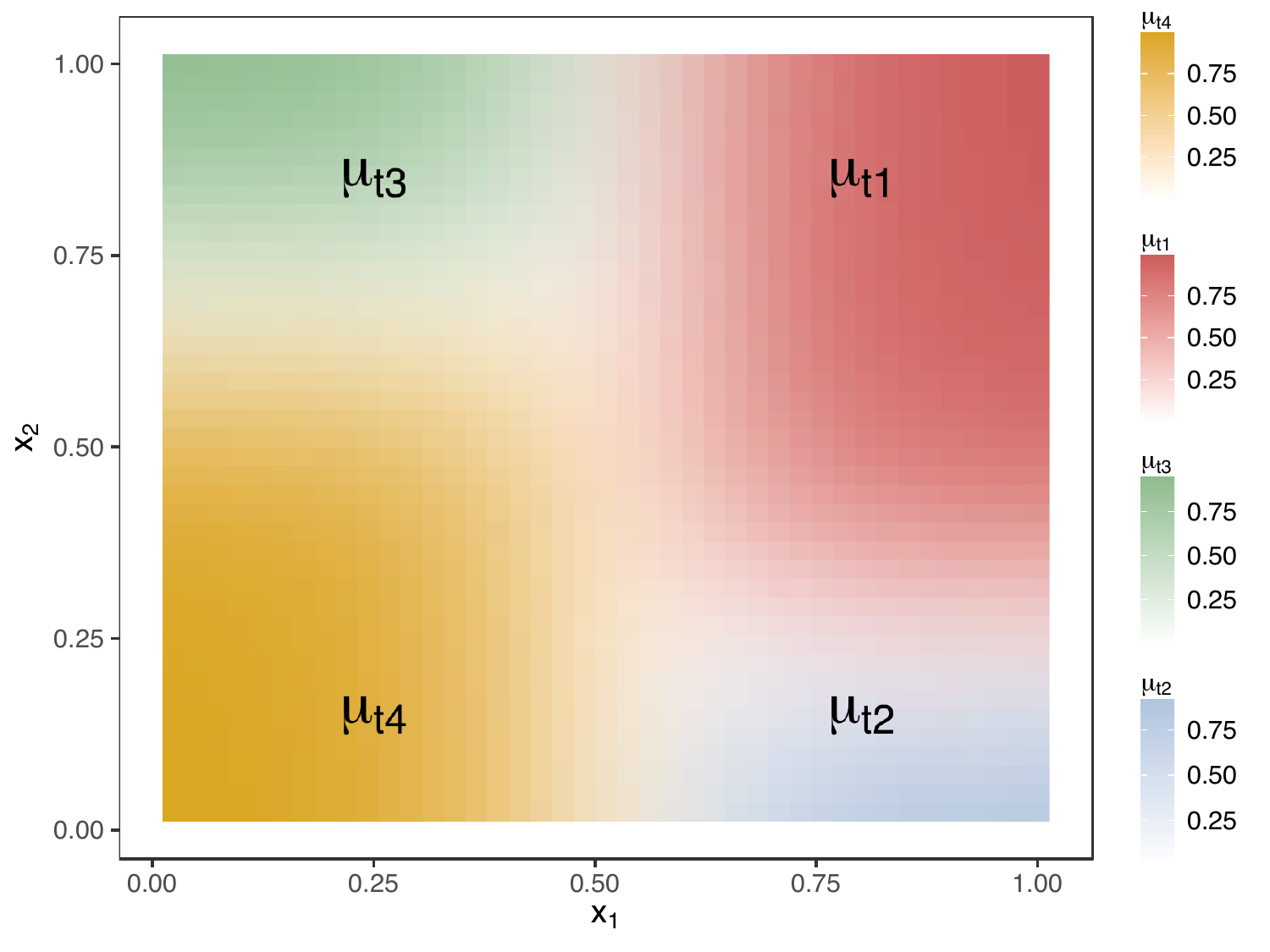} 

}

\caption{Left: a regression tree function correspoding to ``hard''
decision rules. Right: a ``soft'' version of the same regression tree. Points 
are colored according to the respective values of $\phi_{\ell}(x; \mathcal T)$.}\label{fig:softtreefig}
\end{figure}

The \texttt{SoftBart} model modifies the sum of decision trees \(r(x)\)
by replacing the regression trees
\(\operatorname{Tree}(x; \mathcal T_t, \mathcal M_t)\) with \emph{soft}
regression trees \citep{irsoy2012soft}. To generalize a regression tree
to a soft regression tree, we begin by noting that we can rewrite \[
  \operatorname{Tree}(x; \mathcal T, \mathcal M) 
    = \sum_{\ell \in \mathcal L(\mathcal T)} \mu_\ell \, \phi_\ell(x; \mathcal T)
\] where \(\phi_\ell(x; \mathcal T)\) is the indicator function of the
event that \(x\) is associated to leaf \(\ell\) of tree \(\mathcal T\).
Notice that we can rewrite \(\phi_\ell(x; \mathcal T)\) in terms of the
branch rules as \begin{align}
  \label{eq:treeweights}
  \phi_\ell(x; \mathcal T) = \prod_{b \in A(\ell)} 
    I(x_{j_b} \le C_b)^{L_b} \, I(x_{j_b} > C_b)^{1 - L_b},
\end{align} where \(A(\ell)\) is the set of leaf nodes that are
\emph{ancestral} to leaf node \(\ell\) and \(L_b = 1\) if the path from
the root to \(\ell\) goes left at \(b\) and \(L_b = 0\) if the path goes
right; for example, the leaf with \(\mu_3\) in Figure \ref{fig:treefig}
has \(A(\ell)\) consisting only of the root and has
\(L_{\text{root}} = 0\) since the path from the root goes right rather
than left. The indicator functions in \eqref{eq:treeweights} are not
ideal, as they encode a sharp jump from \(0\) to \(1\) as we increase
\(x_b\) from \(x_b \le C_b\) to \(x_b > C_b\).
\citet{linero2017abayesian} generalize \eqref{eq:treeweights} by
replacing the ``hard'' decision rules \(I(x_{j} \le C)\) with \emph{soft
decision rules} \(\psi\left(\frac{x_j - C}{\tau}\right)\), where
\(\psi(x)\) is the cumulative distribution function of a symmetric
random variable. The modified weights become \begin{align}
  \label{eq:softweights}
  \phi_\ell(x; \mathcal T) 
  = \prod_{b \in A(\ell)} \psi\left(\frac{x_{j_b} - C_b}{\tau}\right)^{L_b}
    \left\{1 - \psi\left(\frac{x_{j_b} - C_b}{\tau}\right)\right\}^{1 - L_b}.
\end{align} Because the function \(\psi(x)\) is continuous, the soft
decision tree is continuous in \(x\). The parameter \(\tau\) controls
how ``sharp'' the decisions are, with the limit \(\tau \to 0\)
corresponding to a standard decision tree. Figure \ref{fig:softtreefig}
compares a hard decision tree to a soft decision tree in terms of the
associated functions; we see that the soft decision tree smooths over
the decision boundaries of the hard decision tree. \texttt{SoftBart}
uses the logistic function \(\psi(x) = (1 + \exp(-x))^{-1}\) and gives
each tree \(\mathcal T_t\) its own bandwidth parameter \(\tau_t\) with
\(\tau_t \sim \operatorname{Exponential}(\texttt{scale} = 0.1)\).

\hypertarget{scaling-the-outcome-and-covariates}{%
\subsection{Scaling the Outcome and
Covariates}\label{scaling-the-outcome-and-covariates}}

In order to ensure that the default prior used by \texttt{SoftBart} is
widely appropriate, most functions in the package either work with, or
assume, that a default scaling has been used. The covariates \(X_{ij}\)
are assumed to have been scaled to lie in the interval \([0,1]\). The
model fitting functions (\texttt{softbart},
\texttt{softbart\_regression()}, \texttt{softbart\_probit()}, etc)
perform this standardization automatically by applying a \emph{quantile
transformation} to each numeric covariate, i.e., each covariate is
passed through its empirical cdf, and categorical variables with \(C\)
levels are binarized by introducing \(C\) ``dummy'' variables indicating
the factor level. I generally recommend avoiding the legacy function
\texttt{softbart()}, particularly when working with categorical
variables, as this requires the user to manually preprocess the data.

Models with a continuous outcome \(Y_i\) preprocess the outcome as well.
The \texttt{softbart()} function applies a linear transformation to put
\(Y_i\) in the interval \([-0.5, 0.5]\) as recommended by
\citet{chipman2010bart}, while later functions simply standardize
\(Y_i\) to have mean \(0\) and variance \(1\). I have not found the
choice of standardization for \(Y_i\) to have a large impact on the
results.

Fitting custom models using the \texttt{MakeForest()} functionality
described in Section \ref{embedding-softbart-into-other-models}
unfortunately requires users to scale the covariates and outcomes on
their own, and it is also recommended that users think carefully about
the hyperparameters they use, as there is no guarantee that what is
appropriate for regression/classification problems is appropriate in
general.

\hypertarget{fitting-softbart-models}{%
\subsection{Fitting SoftBart Models}\label{fitting-softbart-models}}

\texttt{SoftBart} models are fit using a Markov chain Monte Carlo
algorithm referred to as \emph{Bayesian backfitting} to approximately
sample realizations of \(r(x)\) from the posterior distribution. For
more details, see \citet{kapelner2014bartmachine} or
\citet{linero2017abayesian}, and for a comprehensive treatment of Markov
chain Monte Carlo see \citet{robert2004monte}. Bayesian backfitting for
the semiparametric Gaussian model proceeds by defining the residuals
\(R_{it} = Y_i - \sum_{k \ne t} \operatorname{Tree}(X_i ; \mathcal T_k, \mathcal M_k)\)
and then noting that
\(R_{it} \sim \operatorname{Normal}\{\operatorname{Tree}(X_i; \mathcal T_t, \mathcal M_t), \sigma^2\}\).
A valid Gibbs sampler could then proceed by iteratively sampling
\((\mathcal T_t, \mathcal M_t)\) from the posterior distribution of the
single-tree model with conditional \begin{align*}
  \pi(\mathcal T_t, \mathcal M_t \mid \text{Data}, \mathcal T_{-t}, \mathcal M_{-t})
  \propto
  \prod_i \operatorname{Normal}\{R_{it} \mid \operatorname{Tree}(X_i; \mathcal T_t, \mathcal M_t), \sigma^2\} \, 
      \pi_\mathcal T(\mathcal T_t) \, \pi_{\mathcal M}(\mathcal M_t \mid \mathcal T_t)
\end{align*} This can be sampled via Markov chain Monte Carlo in the
following steps:

\begin{enumerate}
\def\labelenumi{\arabic{enumi}.}
\item
  Integrate out \(\mathcal M_t\) to obtain the marginal likelhood
  \(L(\mathcal T_t) = \pi_{\mathcal T}(\mathcal T_t) \int \operatorname{Normal}\{R_{it} \mid \operatorname{Tree}(X_i; \mathcal T_t, \mathcal M_t), \sigma^2\} \, \pi_{\mathcal M}(\mathcal M\mid \mathcal T_t) \ d\mathcal M\).
\item
  Propose a modification \(\mathcal T' \sim q(\cdot \mid \mathcal T_t)\)
  to \(\mathcal T_t\) and accept this modification with probability
  \(\frac{L(\mathcal T') \, q(\mathcal T_t \mid \mathcal T')}{L(\mathcal T_t) \, q(\mathcal T' \mid \mathcal T_t)} \wedge 1\)
  (otherwise, leave \(\mathcal T_t\) unchanged).
\item
  Sample \(\mathcal M_t\) from its full conditional given
  \(\mathcal T_t\).
\end{enumerate}

\citet{linero2017abayesian} shows how to carry out these individual
steps in the case of \texttt{SoftBart}. As we show, this framework can
be extended to many other models of interest, with these steps being
automatically carried out in custom \texttt{SoftBart} models.

\hypertarget{softbart-in-action}{%
\section{SoftBart in Action}\label{softbart-in-action}}

The \texttt{SoftBart} package is available on CRAN and can be installed
by running

\begin{Shaded}
\begin{Highlighting}[]
\FunctionTok{install.packages}\NormalTok{(}\StringTok{"SoftBart"}\NormalTok{)}
\end{Highlighting}
\end{Shaded}

Alternatively, for the most up-to-date version of the software,
\texttt{SoftBart} can be installed from source using the
\texttt{devtools} package:

\begin{Shaded}
\begin{Highlighting}[]
\NormalTok{devtools}\SpecialCharTok{::}\FunctionTok{install\_github}\NormalTok{(}\StringTok{"www.github.com/theodds/SoftBart"}\NormalTok{)}
\end{Highlighting}
\end{Shaded}

I show how to use the \texttt{softbart} and
\texttt{softbart\_regression} functions to fit a semiparametric Gaussian
regression model. Both of these functions fit the same model; the
difference is that \texttt{softbart} is designed to mirror the usage of
the \texttt{bart} function in the original \texttt{BayesTree} package,
while \texttt{softbart\_regression} specifies models using formulas and
also has an associated \texttt{predict} function for predicting on data
that was not passed to the function.

\hypertarget{the-softbart-function}{%
\subsection{The softbart Function}\label{the-softbart-function}}

I first illustrate the \texttt{softbart} function, which has users pass
a matrix of covariates \texttt{X}, a vector of outcomes \texttt{Y}, and
a test set of covariates \texttt{X\_test} that the user wants to predict
\(r(x)\) at. Below we generate data under a simulation setting of
\citet{friedman1991multivariate}, which takes \begin{align}
  \label{eq:friedman}
  Y_i \sim \operatorname{Normal}\{r_0(X_i), \sigma_0^2\}
  \quad \text{where} \quad
  r_0(x) = 10 \sin(\pi x_1 x_2) + 20(x_3 - 0.5)^2 + 10 x_4 + 5 x_5,
\end{align} where \(\sigma_0 = 1\) and
\(X_i \stackrel{\text{iid}}{\sim}\operatorname{Uniform}([0,1]^P)\). Note
that only \(X_{i1},\ldots,X_{i5}\) are relevant, and \(X_{ij}\) is a
``nuisance predictor'' when \(j > 5\). We then use the function
\texttt{softbart} to fit the semiparametric Gaussian regression model.

\begin{Shaded}
\begin{Highlighting}[]
\FunctionTok{set.seed}\NormalTok{(}\DecValTok{1212}\NormalTok{)}
\NormalTok{sim\_fried }\OtherTok{\textless{}{-}} \ControlFlowTok{function}\NormalTok{(N,P,sigma) \{}
\NormalTok{  X  }\OtherTok{\textless{}{-}} \FunctionTok{matrix}\NormalTok{(}\FunctionTok{runif}\NormalTok{(N }\SpecialCharTok{*}\NormalTok{ P), }\AttributeTok{nrow =}\NormalTok{ N, }\AttributeTok{ncol =}\NormalTok{ P)}
\NormalTok{  mu }\OtherTok{\textless{}{-}} \DecValTok{10} \SpecialCharTok{*} \FunctionTok{sin}\NormalTok{(pi }\SpecialCharTok{*}\NormalTok{ X[,}\DecValTok{1}\NormalTok{] }\SpecialCharTok{*}\NormalTok{ X[,}\DecValTok{2}\NormalTok{]) }\SpecialCharTok{+} \DecValTok{20} \SpecialCharTok{*}\NormalTok{ (X[,}\DecValTok{3}\NormalTok{] }\SpecialCharTok{{-}} \FloatTok{0.5}\NormalTok{)}\SpecialCharTok{\^{}}\DecValTok{2} \SpecialCharTok{+} 
    \DecValTok{10} \SpecialCharTok{*}\NormalTok{ X[,}\DecValTok{4}\NormalTok{] }\SpecialCharTok{+} \DecValTok{5} \SpecialCharTok{*}\NormalTok{ X[,}\DecValTok{5}\NormalTok{]}
\NormalTok{  Y }\OtherTok{\textless{}{-}}\NormalTok{ mu }\SpecialCharTok{+}\NormalTok{ sigma }\SpecialCharTok{*} \FunctionTok{rnorm}\NormalTok{(N)}
  \FunctionTok{return}\NormalTok{(}\FunctionTok{data.frame}\NormalTok{(}\AttributeTok{X =}\NormalTok{ X, }\AttributeTok{Y =}\NormalTok{ Y, }\AttributeTok{mu =}\NormalTok{ mu))}
\NormalTok{\}}

\NormalTok{training\_data }\OtherTok{\textless{}{-}} \FunctionTok{sim\_fried}\NormalTok{(}\DecValTok{250}\NormalTok{, }\DecValTok{250}\NormalTok{, }\DecValTok{1}\NormalTok{)}
\NormalTok{test\_data     }\OtherTok{\textless{}{-}} \FunctionTok{sim\_fried}\NormalTok{(}\DecValTok{250}\NormalTok{, }\DecValTok{250}\NormalTok{, }\DecValTok{1}\NormalTok{)}

\NormalTok{X\_train }\OtherTok{\textless{}{-}} \FunctionTok{model.matrix}\NormalTok{(Y }\SpecialCharTok{\textasciitilde{}}\NormalTok{ . }\SpecialCharTok{{-}} \DecValTok{1} \SpecialCharTok{{-}}\NormalTok{ mu, }\AttributeTok{data =}\NormalTok{ training\_data)}
\NormalTok{X\_test  }\OtherTok{\textless{}{-}} \FunctionTok{model.matrix}\NormalTok{(Y }\SpecialCharTok{\textasciitilde{}}\NormalTok{ . }\SpecialCharTok{{-}} \DecValTok{1} \SpecialCharTok{{-}}\NormalTok{ mu, }\AttributeTok{data =}\NormalTok{ test\_data)}

\NormalTok{fitted\_softbart }\OtherTok{\textless{}{-}} \FunctionTok{softbart}\NormalTok{(}\AttributeTok{X =}\NormalTok{ X\_train, }\AttributeTok{Y =}\NormalTok{ training\_data}\SpecialCharTok{$}\NormalTok{Y, }
                            \AttributeTok{X\_test =}\NormalTok{ X\_test)}
\end{Highlighting}
\end{Shaded}

The \texttt{softbart} function returns an object of type
\texttt{softbart}, and the associated \texttt{plot} function displays a
traceplot for the parameter \(\sigma\) and a plot of the outcome \(Y_i\)
against its prediction \(\widehat r(X_i)\) where \(\widehat r(x)\) is
the posterior mean of \(r(x)\).

\begin{Shaded}
\begin{Highlighting}[]
\FunctionTok{plot}\NormalTok{(fitted\_softbart)}
\end{Highlighting}
\end{Shaded}

\begin{figure}[t]

{\centering \includegraphics{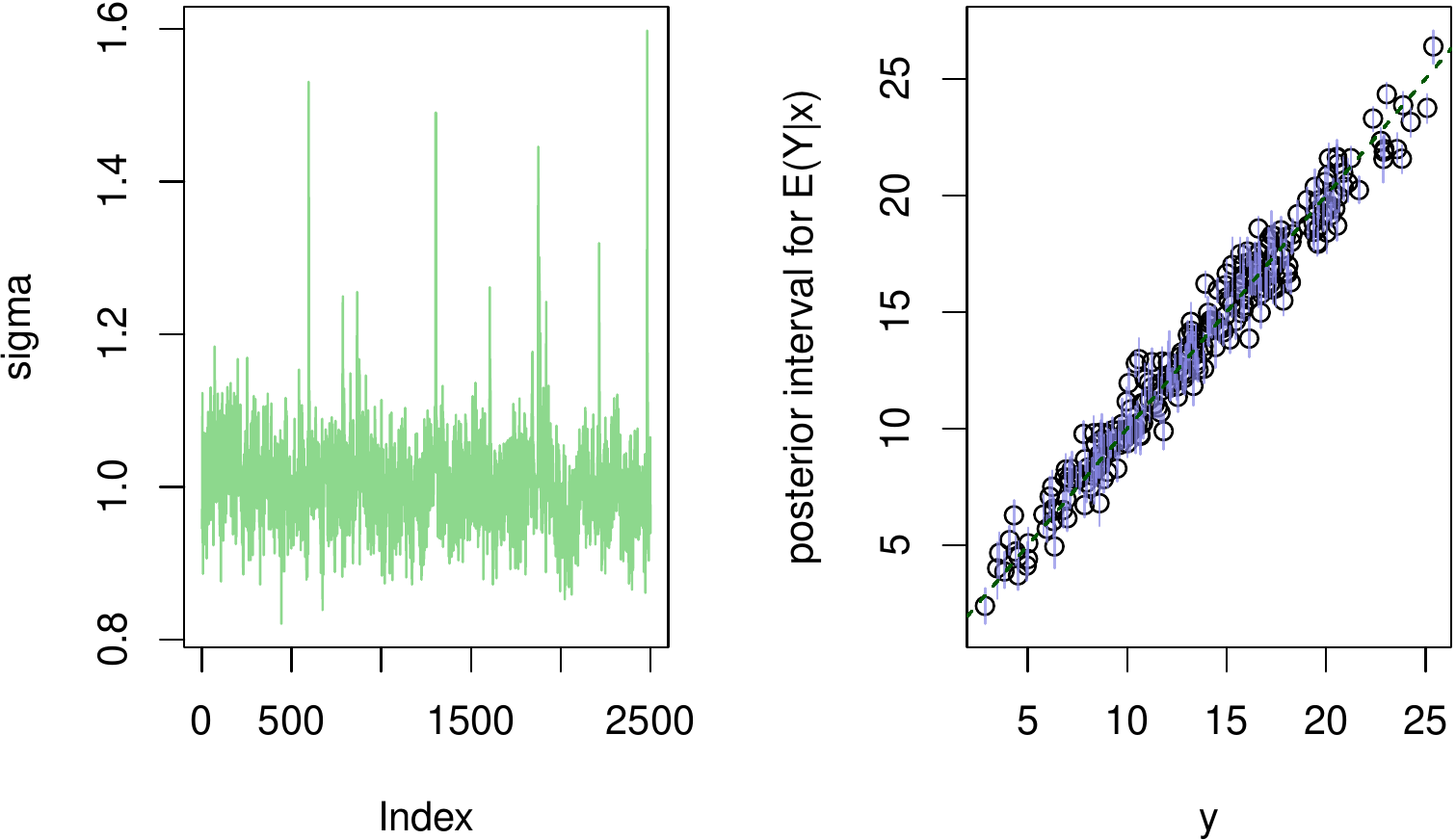} 

}

\caption{Left: traceplot of $\sigma$ for the model fit using \texttt{softbart}. Right: plot of the predictions against their posterior predicted values, with vertical lines representing 95\% confidence (not prediction) intervals for the mean outcome.}\label{fig:plotfit}
\end{figure}

The output is given in Figure \ref{fig:plotfit}, and we see that the
posterior distribution of \(\sigma\) is concentrated around its true
value \(\sigma_0 = 1\), the chain for \(\sigma\) mixes quite well, and
the predictions produced by \texttt{softbart()} are quite close to their
true values. We can also check the prediction error on the test set
using the \texttt{rmse} function, which returns
\(\texttt{rmse(x,y)} = \sqrt{\frac{1}{N} \sum_i (\texttt{x}_i - \texttt{y}_i)^2}\).
The Bayes estimates of \(r_0(X_i)\) for the training and test sets are
given by \texttt{y\_hat\_train\_mean} and \texttt{y\_hat\_test\_mean},
respectively. We compare the true values of \(r_0(X_i)\) to their
estimates as:

\begin{Shaded}
\begin{Highlighting}[]
\FunctionTok{rmse}\NormalTok{(fitted\_softbart}\SpecialCharTok{$}\NormalTok{y\_hat\_train\_mean, training\_data}\SpecialCharTok{$}\NormalTok{mu)}
\end{Highlighting}
\end{Shaded}

\begin{verbatim}
## [1] 0.4172506
\end{verbatim}

\begin{Shaded}
\begin{Highlighting}[]
\FunctionTok{rmse}\NormalTok{(fitted\_softbart}\SpecialCharTok{$}\NormalTok{y\_hat\_test\_mean, test\_data}\SpecialCharTok{$}\NormalTok{mu)}
\end{Highlighting}
\end{Shaded}

\begin{verbatim}
## [1] 0.4404536
\end{verbatim}

The raw samples of \(r(X_i)\) for the training and test sets are given
by \texttt{y\_hat\_train} and \texttt{y\_hat\_test}, which are matrices
with rows corresponding to samples from the Markov chain. These can be
used to, among other things, construct credible intervals for the
predicted values from the model. For example, the following code
constructs a credible interval for \(r(X_1)\):

\begin{Shaded}
\begin{Highlighting}[]
\FunctionTok{quantile}\NormalTok{(fitted\_softbart}\SpecialCharTok{$}\NormalTok{y\_hat\_train[,}\DecValTok{1}\NormalTok{], }\FunctionTok{c}\NormalTok{(}\FloatTok{0.025}\NormalTok{, }\FloatTok{0.975}\NormalTok{))}
\end{Highlighting}
\end{Shaded}

\begin{verbatim}
##     2.5%    97.5% 
## 18.82388 20.24865
\end{verbatim}

which contains the true value \texttt{training\_data\$mu{[}1{]}} = 19.3.

\hypertarget{model-options-hypers-and-opts}{%
\subsection{Model Options: Hypers and
Opts}\label{model-options-hypers-and-opts}}

Users may also wish to customize the model hyperparameters or change the
behavior of the Markov chain (for example, they may want to collect more
samples, fix some hyperparameters to be constant, or thin the Markov
chain). This can be done by setting the \texttt{hypers} and
\texttt{opts} arguments of \texttt{softbart}. These arguments can be
constructed using the functions \texttt{Hypers()} and \texttt{Opts()},
respectively. For example, the following code uses \texttt{Opts()} to
construct an \texttt{opts} argument that (i) increases the number of
samples, (ii) turns off the updating of the splitting proportion vector
\(s\), and (iii) turns off updating of \(\sigma_\mu\):

\begin{Shaded}
\begin{Highlighting}[]
\NormalTok{opts }\OtherTok{\textless{}{-}} \FunctionTok{Opts}\NormalTok{(}\AttributeTok{num\_burn =} \DecValTok{5000}\NormalTok{, }\AttributeTok{num\_save =} \DecValTok{5000}\NormalTok{, }
             \AttributeTok{update\_s =} \ConstantTok{FALSE}\NormalTok{, }\AttributeTok{update\_sigma\_mu =} \ConstantTok{FALSE}\NormalTok{)}
\end{Highlighting}
\end{Shaded}

The \texttt{Hypers()} function can similarly be used to change the
hyperparameters. For example, we can modify the tree-growing prior by
changing the values of \(\gamma\) and \(\beta\) in
\eqref{eq:tree-split-prob} and increase the number of trees \(T\) as
follows:

\begin{Shaded}
\begin{Highlighting}[]
\NormalTok{hypers }\OtherTok{\textless{}{-}} \FunctionTok{Hypers}\NormalTok{(}\AttributeTok{X =}\NormalTok{ X\_train, }\AttributeTok{Y =}\NormalTok{ training\_data}\SpecialCharTok{$}\NormalTok{Y, }
                 \AttributeTok{num\_tree =} \DecValTok{50}\NormalTok{, }\AttributeTok{beta =} \DecValTok{1}\NormalTok{, }\AttributeTok{gamma =} \FloatTok{0.9}\NormalTok{)}
\end{Highlighting}
\end{Shaded}

The objects \texttt{hypers} and \texttt{opts} can then be passed to the
function \texttt{softbart()}.

\begin{Shaded}
\begin{Highlighting}[]
\FunctionTok{set.seed}\NormalTok{(}\DecValTok{19320}\NormalTok{)}
\NormalTok{fitted\_softbart\_2 }\OtherTok{\textless{}{-}} \FunctionTok{softbart}\NormalTok{(X\_train, training\_data}\SpecialCharTok{$}\NormalTok{Y, X\_test, }
                              \AttributeTok{opts =}\NormalTok{ opts, }\AttributeTok{hypers =}\NormalTok{ hypers)}
\end{Highlighting}
\end{Shaded}

We can then check the performance of the model:

\begin{Shaded}
\begin{Highlighting}[]
\FunctionTok{rmse}\NormalTok{(test\_data}\SpecialCharTok{$}\NormalTok{mu, fitted\_softbart\_2}\SpecialCharTok{$}\NormalTok{y\_hat\_test\_mean)}
\end{Highlighting}
\end{Shaded}

\begin{verbatim}
## [1] 1.175639
\end{verbatim}

We see that this new model performs worse than the old one; this is
because the ground truth \(r_0(x)\) is sparse, but we have turned off
the variable selection prior for \(s\).

\texttt{Hypers()} and \texttt{Opts()} can be used to modify many other
settings, and this is often required when embedding \texttt{SoftBart}
into other models. We revisit the usage of \texttt{Hypers()} and
\texttt{Opts()} for this purpose in Section
\ref{embedding-softbart-into-other-models}.

\hypertarget{other-options-for-fitting-models}{%
\subsection{Other Options for Fitting
Models}\label{other-options-for-fitting-models}}

The \texttt{softbart()} function is designed to match the
\texttt{BayesTree} package in terms of usage. Other functions in
\texttt{SoftBart} instead use model specifications that are in line with
how users specify (say) linear models using \texttt{lm()}. For example,
the \texttt{softbart\_regression()} function allows users to specify a
model using a formula:

\begin{Shaded}
\begin{Highlighting}[]
\NormalTok{fitted\_regression }\OtherTok{\textless{}{-}} \FunctionTok{softbart\_regression}\NormalTok{(Y }\SpecialCharTok{\textasciitilde{}}\NormalTok{ . }\SpecialCharTok{{-}}\NormalTok{ mu, }
                                         \AttributeTok{data =}\NormalTok{ training\_data, }
                                         \AttributeTok{test\_data =}\NormalTok{ test\_data)}
\end{Highlighting}
\end{Shaded}

The \texttt{softbart\_regression()} function returns an object of type
\texttt{softbart\_regression}, which also has several advantages over
the output of \texttt{softbart()}. For example,
\texttt{softbart\_regression()} has an associated \texttt{predict()}
generic that can be used to predict on data after the model is fit.
Additionally, \texttt{softbart\_regression()} and the other functions I
discuss (\texttt{softbart\_probit()}, \texttt{gsoftbart\_regression()}
and \texttt{vc\_softbart\_regression()}) are designed to work with data
frames rather than matrices and allow for factors to be passed directly
as predictors.

Both the \texttt{softbart\_regression()} and \texttt{softbart\_probit()}
functions produce objects that can be used with the \texttt{predict()}
generic, provided \texttt{Opts()} is called with
\texttt{cache\_trees\ =\ TRUE}, which is done by default. We predict as
follows:

\begin{Shaded}
\begin{Highlighting}[]
\NormalTok{predicted\_values }\OtherTok{\textless{}{-}} \FunctionTok{predict}\NormalTok{(fitted\_regression, test\_data)}
\FunctionTok{names}\NormalTok{(predicted\_values)}
\end{Highlighting}
\end{Shaded}

\begin{verbatim}
## [1] "mu"      "mu_mean"
\end{verbatim}

For \texttt{softbart\_regression()}, \texttt{predict()} returns both
samples of the predicted values (\texttt{mu}) on the test set and their
posterior means (\texttt{mu\_mean}).

\hypertarget{partial-dependence-plots}{%
\subsection{Partial Dependence Plots}\label{partial-dependence-plots}}

A common method for summarizing the information contained in black-box
models is to construct a \emph{partial dependence plot}
\citep{friedman1991multivariate}. A partial dependence plot for
predictor \(j\) is constructed from a \emph{partial dependence function}
\begin{align*}
  \operatorname{PD}_j(v) 
    = \frac{1}{N} 
    \sum_{i=1}^N r(X_{i1}, \ldots, X_{i(j-1)}, v, X_{i(j+1)}, \ldots, X_{iP}).
\end{align*} This reduces a multivariate function \(r(x)\) to a
univariate function that is much easier to visualize. Like other BART
packages, \texttt{SoftBart} makes it easy to compute samples of
\(\operatorname{PD}_j(v)\) at a specified collection of \(v\)'s. This
can be done using the \texttt{partial\_dependence\_regression()}
function on \texttt{softbart\_regression} objects. For example, using
our model fit to \eqref{eq:friedman} we can estimate
\(\operatorname{PD}_4\) as

\begin{Shaded}
\begin{Highlighting}[]
\NormalTok{grid\_x4 }\OtherTok{\textless{}{-}} \FunctionTok{seq}\NormalTok{(}\AttributeTok{from =} \DecValTok{0}\NormalTok{, }\AttributeTok{to =} \DecValTok{1}\NormalTok{, }\AttributeTok{length =} \DecValTok{10}\NormalTok{)}
\NormalTok{pdf\_x4 }\OtherTok{\textless{}{-}} \FunctionTok{partial\_dependence\_regression}\NormalTok{(fitted\_regression, }
\NormalTok{                                        training\_data, }\StringTok{"X.4"}\NormalTok{, grid\_x4)}
\end{Highlighting}
\end{Shaded}

\begin{Shaded}
\begin{Highlighting}[]
\FunctionTok{library}\NormalTok{(tidyverse)}
\NormalTok{pdf\_offset }\OtherTok{\textless{}{-}} \FunctionTok{mean}\NormalTok{(training\_data}\SpecialCharTok{$}\NormalTok{mu }\SpecialCharTok{{-}} \DecValTok{10} \SpecialCharTok{*}\NormalTok{ training\_data}\SpecialCharTok{$}\NormalTok{X}\FloatTok{.4}\NormalTok{)}
\FunctionTok{ggplot}\NormalTok{(pdf\_x4}\SpecialCharTok{$}\NormalTok{pred\_df, }\FunctionTok{aes}\NormalTok{(}\AttributeTok{x =}\NormalTok{ X}\FloatTok{.4}\NormalTok{, }\AttributeTok{y =}\NormalTok{ mu)) }\SpecialCharTok{+} 
  \FunctionTok{geom\_line}\NormalTok{(}\AttributeTok{stat =} \StringTok{"summary"}\NormalTok{, }\AttributeTok{fun =}\NormalTok{ mean) }\SpecialCharTok{+} 
  \FunctionTok{geom\_ribbon}\NormalTok{(}\AttributeTok{stat =} \StringTok{"summary"}\NormalTok{, }\AttributeTok{alpha =} \FloatTok{0.3}\NormalTok{,}
              \AttributeTok{fun.min =} \ControlFlowTok{function}\NormalTok{(x) }\FunctionTok{quantile}\NormalTok{(x, }\FloatTok{0.025}\NormalTok{), }
              \AttributeTok{fun.max =} \ControlFlowTok{function}\NormalTok{(x) }\FunctionTok{quantile}\NormalTok{(x, }\FloatTok{0.975}\NormalTok{)) }\SpecialCharTok{+} 
  \FunctionTok{xlab}\NormalTok{(}\StringTok{"$X\_4$"}\NormalTok{) }\SpecialCharTok{+} \FunctionTok{ylab}\NormalTok{(}\StringTok{"$}\SpecialCharTok{\textbackslash{}\textbackslash{}}\StringTok{mbox\{PD\}\_4$"}\NormalTok{) }\SpecialCharTok{+} 
  \FunctionTok{stat\_function}\NormalTok{(}\AttributeTok{fun =} \ControlFlowTok{function}\NormalTok{(x) pdf\_offset }\SpecialCharTok{+} \DecValTok{10} \SpecialCharTok{*}\NormalTok{ x, }
                \AttributeTok{color =} \StringTok{"\#5F96C2"}\NormalTok{, }\AttributeTok{linetype =} \DecValTok{2}\NormalTok{, }\AttributeTok{size =} \DecValTok{2}\NormalTok{) }\SpecialCharTok{+} 
  \FunctionTok{theme\_bw}\NormalTok{()}
\end{Highlighting}
\end{Shaded}

\begin{figure}[t]

{\centering \includegraphics[width=0.5\textwidth,]{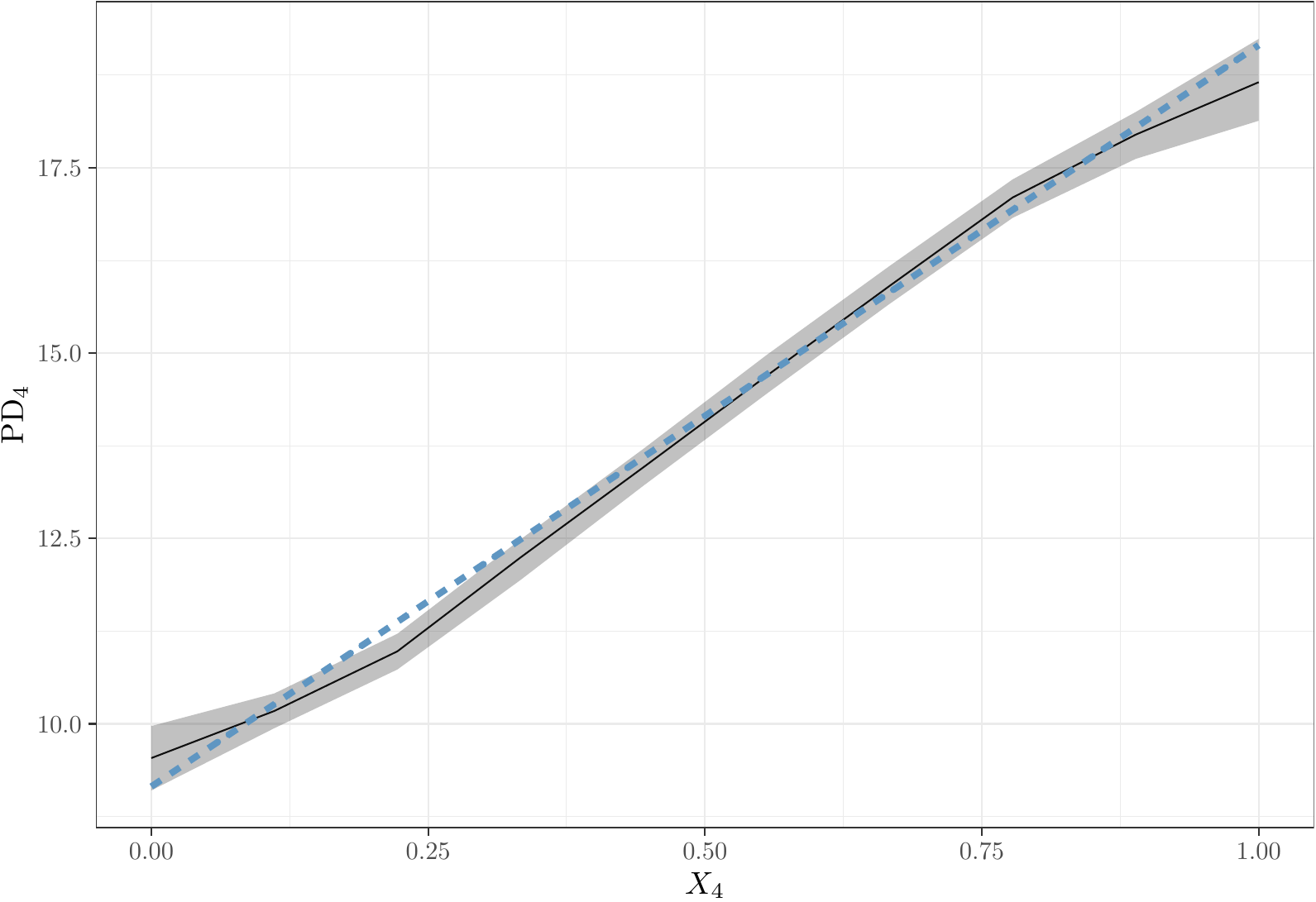} 

}

\caption{Point estimate (solid black line) and credible band (gray) of the partial dependence function for $X_{i4}$, with the true partial dependence function given by the dashed line.}\label{fig:pdpx4}
\end{figure}

For convenience, this function returns both a ``tidy''
\citep{wickham2019welcome} \texttt{data.frame} to be used with
\texttt{ggplot()} \citep{wickham2016ggplot} as well as a matrix of
samples of \(\operatorname{PD}_j(v)\) to be used with base \texttt{R}
plotting functions. Based on the plot (see Figure \ref{fig:pdpx4}), we
see that the fitted model does a good of capturing the true partial
dependence function, which (due to the fact that \(r(x)\) is additive in
\(x_4\)) is equal to \(\operatorname{PD}_4(v) = C + 10v\) for some
constant \(C\).

\hypertarget{variable-selection}{%
\section{Variable Selection}\label{variable-selection}}

By default, \texttt{SoftBart} uses the \emph{sparsity inducing prior}
introduced by \citet{linero2016bayesian} to perform variable selection.
This prior induces sparsity on the vector of splitting proportions by
taking \(s \sim \operatorname{Dirichlet}(\alpha/P, \ldots, \alpha/P)\),
with the idea being that if a variable \(j\) is irrelevant then the
model can remove \(j\) by taking \(s_j\) very small.
\citet{linero2016bayesian} shows that, when both the number of
predictors \(P\) and the number of branches \(B\) in the ensemble are
large, then we have the prior approximation
\(Q - 1 \sim \operatorname{Poisson}(\theta_B)\) where \(Q\) is the
number of relevant predictors and
\(\theta_B = \alpha \sum_{i=0}^{B-1} (\alpha + i)^{-1}\). The value of
\(\alpha\) can be fixed to obtain a desired amount of sparsity a-priori,
but by default \texttt{SoftBart} gives \(\alpha\) a beta-prime
hyperprior
\(\frac{\alpha}{\alpha + P} \sim \operatorname{Beta}(0.5, 1)\).

\begin{figure}[t]

{\centering \includegraphics{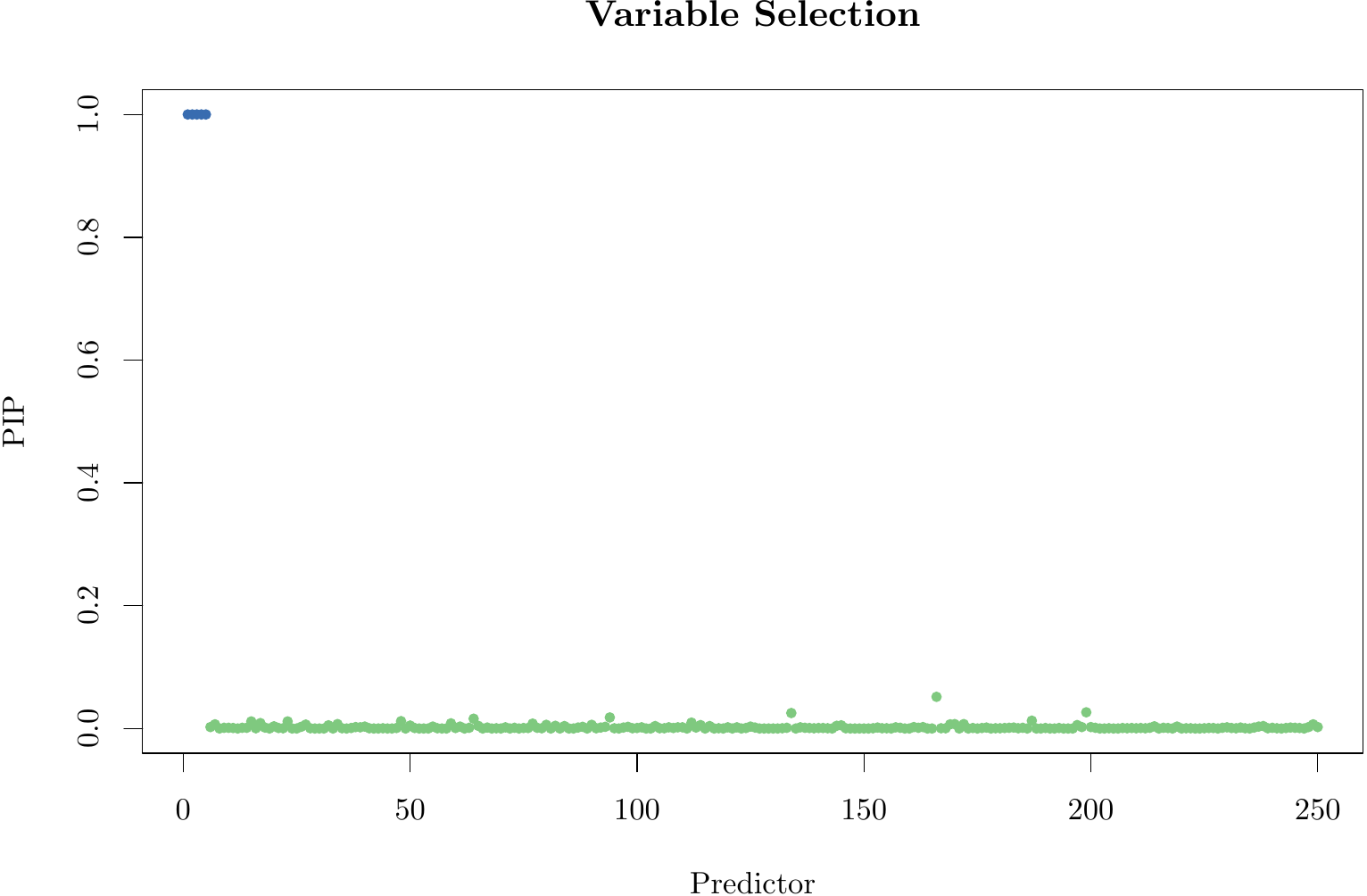} 

}

\caption{Plot of posterior inclusion probabilities produced by \texttt{softbart} using the variable selection prior.}\label{fig:plotpip}
\end{figure}

Variables can be selected according to their \emph{posterior inclusion
probability} \[
  \operatorname{PIP}_j = \Pr(\text{predictor $j$ appears in the ensemble} \mid \text{Data}).
\] We can then extract the \(\operatorname{PIP}\)'s from the model fit
using \texttt{posterior\_probs()} and plot them (Figure
\ref{fig:plotpip}):

\begin{Shaded}
\begin{Highlighting}[]
\NormalTok{variable\_selection }\OtherTok{\textless{}{-}} \FunctionTok{posterior\_probs}\NormalTok{(fitted\_softbart)}
\FunctionTok{plot}\NormalTok{(variable\_selection}\SpecialCharTok{$}\NormalTok{post\_probs, }
     \AttributeTok{col =} \FunctionTok{ifelse}\NormalTok{(}\DecValTok{1}\SpecialCharTok{:}\DecValTok{250} \SpecialCharTok{\textless{}} \DecValTok{6}\NormalTok{, }\StringTok{"\#386CB0"}\NormalTok{, }\StringTok{"\#7FC97F"}\NormalTok{), }\AttributeTok{pch =} \DecValTok{20}\NormalTok{,}
     \AttributeTok{xlab =} \StringTok{"Predictor"}\NormalTok{, }\AttributeTok{ylab =} \StringTok{"PIP"}\NormalTok{, }\AttributeTok{main =} \StringTok{"Variable Selection"}\NormalTok{)}
\end{Highlighting}
\end{Shaded}

The \texttt{posterior\_probs()} function also returns the \emph{median
probability model}, defined by
\(\mathcal S = \{j: \operatorname{PIP}_j \ge 0.5\}\). For the fit to the
\eqref{eq:friedman} data, we see that the median probability model
coincides with the true data generating process:

\begin{Shaded}
\begin{Highlighting}[]
\FunctionTok{print}\NormalTok{(variable\_selection}\SpecialCharTok{$}\NormalTok{median\_probability\_model)}
\end{Highlighting}
\end{Shaded}

\begin{verbatim}
## [1] 1 2 3 4 5
\end{verbatim}

\begin{figure}[t]

{\centering \includegraphics{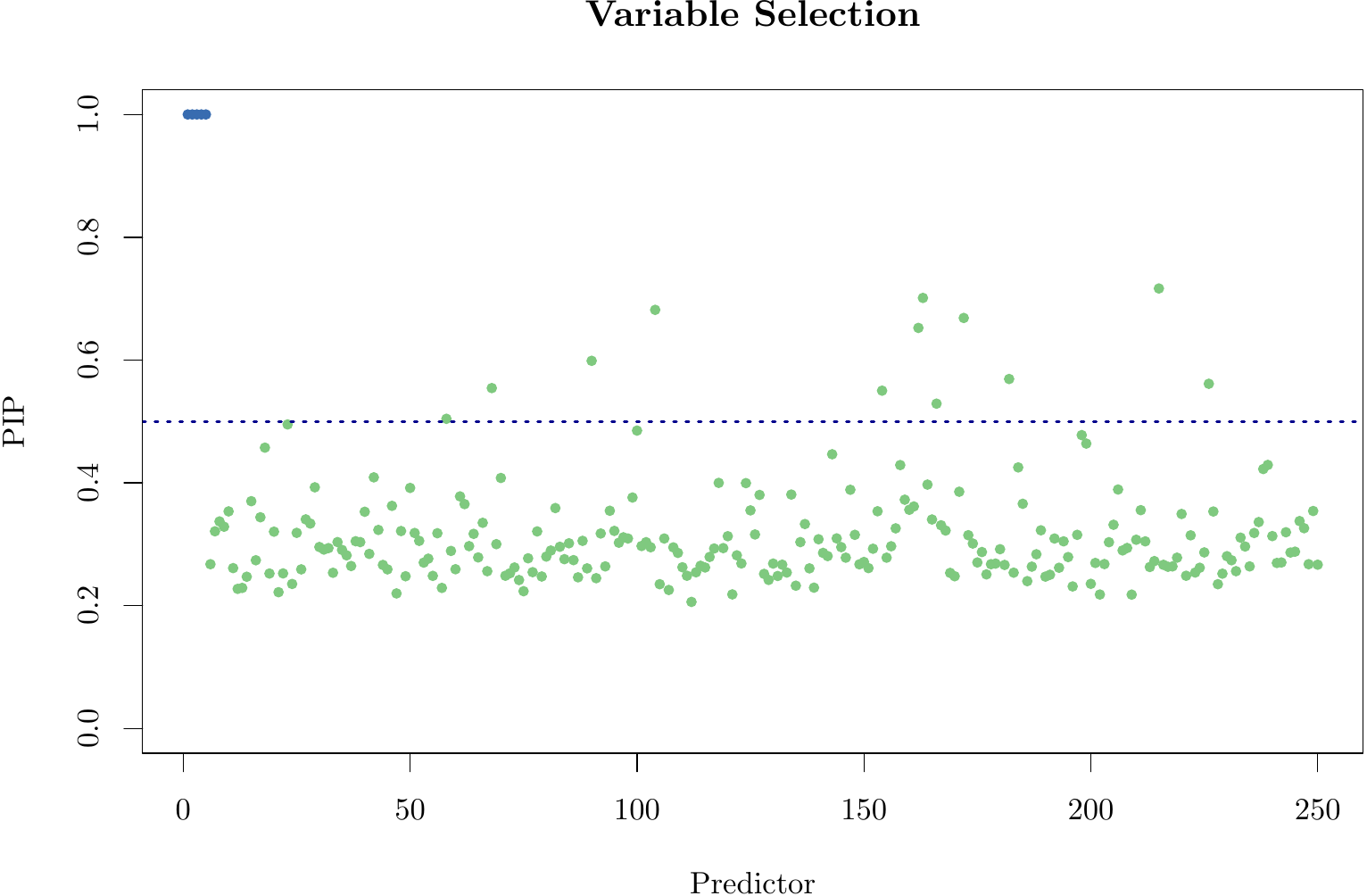} 

}

\caption{Plot of posterior inclusion probabilities produced by \texttt{softbart} without using the variable selection prior.}\label{fig:plotpip2}
\end{figure}

The variable selection prior can be turned off by setting
\texttt{update\_s\ =\ FALSE} in \texttt{Opts()}. For example, we can
compare the above fit to the fit \texttt{fitted\_softbart\_2} (Figure
\ref{fig:plotpip2}), which we recall does not place a prior on \(s\):

\begin{Shaded}
\begin{Highlighting}[]
\NormalTok{variable\_selection\_2 }\OtherTok{\textless{}{-}} \FunctionTok{posterior\_probs}\NormalTok{(fitted\_softbart\_2)}
\FunctionTok{plot}\NormalTok{(variable\_selection\_2}\SpecialCharTok{$}\NormalTok{post\_probs, }
     \AttributeTok{col =} \FunctionTok{ifelse}\NormalTok{(}\DecValTok{1}\SpecialCharTok{:}\DecValTok{250} \SpecialCharTok{\textless{}} \DecValTok{6}\NormalTok{, }\StringTok{"\#386CB0"}\NormalTok{, }\StringTok{"\#7FC97F"}\NormalTok{), }\AttributeTok{pch =} \DecValTok{20}\NormalTok{,}
     \AttributeTok{xlab =} \StringTok{"Predictor"}\NormalTok{, }\AttributeTok{ylab =} \StringTok{"PIP"}\NormalTok{, }\AttributeTok{main =} \StringTok{"Variable Selection"}\NormalTok{, }
     \AttributeTok{ylim =} \FunctionTok{c}\NormalTok{(}\DecValTok{0}\NormalTok{,}\DecValTok{1}\NormalTok{))}
\FunctionTok{abline}\NormalTok{(}\AttributeTok{h =} \FloatTok{0.5}\NormalTok{, }\AttributeTok{col =} \StringTok{"darkblue"}\NormalTok{, }\AttributeTok{lwd =} \DecValTok{2}\NormalTok{, }\AttributeTok{lty =} \DecValTok{3}\NormalTok{)}
\end{Highlighting}
\end{Shaded}

We see that there are many more variables selected by the median
probability model when the variable selection prior is not used and,
moreover, all of the predictors have a \(\operatorname{PIP}\) higher
than 0.2. For this reason, some works
\citep{chipman2010bart, bleich2014bayesian} recommend using only a small
number of trees when the end goal is variable selection, as this forces
the variables to ``compete'' for splitting rules in the ensemble. The
use of the variable selection prior makes this restriction to small
numbers of trees largely unnecessary.

\begin{figure}[t]

{\centering \includegraphics{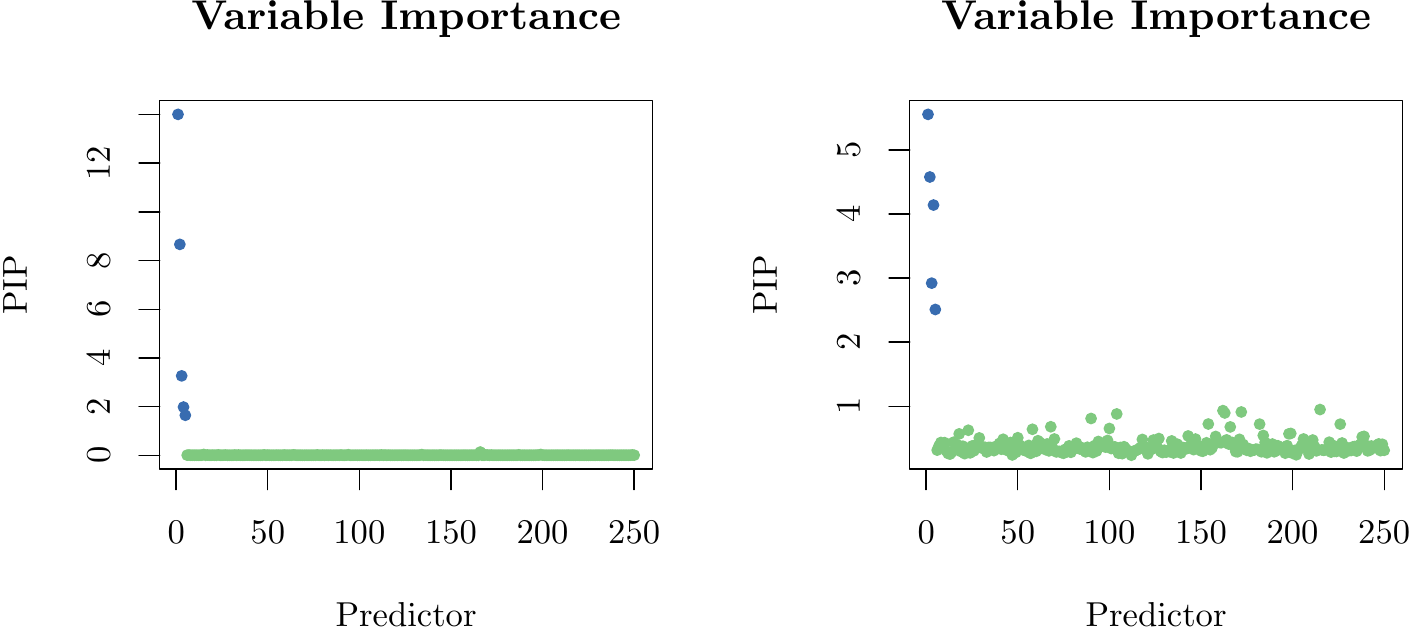} 

}

\caption{Left: variable importance when using the variable selection prior for $s$. Right: variable importances without using the variable selection prior for $s$.}\label{fig:plotvarimp}
\end{figure}

\citet{chipman2010bart} also suggest using the number of times a
predictor is used in the ensemble as a measure of overall variable
importance; this idea, or ideas like it, have also been used to measure
the importance of variables for other types of tree-based machine
learning algorithms, such as random forests
\citep{breiman2001random, diaz2006gene}. The variable importances are
given by the quantity \texttt{varimp} (see Figure \ref{fig:plotvarimp}):

\begin{Shaded}
\begin{Highlighting}[]
\FunctionTok{par}\NormalTok{(}\AttributeTok{mfrow =} \FunctionTok{c}\NormalTok{(}\DecValTok{1}\NormalTok{,}\DecValTok{2}\NormalTok{))}
\FunctionTok{plot}\NormalTok{(variable\_selection}\SpecialCharTok{$}\NormalTok{varimp, }
     \AttributeTok{col =} \FunctionTok{ifelse}\NormalTok{(}\DecValTok{1}\SpecialCharTok{:}\DecValTok{250} \SpecialCharTok{\textless{}} \DecValTok{6}\NormalTok{, }\StringTok{"\#386CB0"}\NormalTok{, }\StringTok{"\#7FC97F"}\NormalTok{), }\AttributeTok{pch =} \DecValTok{20}\NormalTok{,}
     \AttributeTok{xlab =} \StringTok{"Predictor"}\NormalTok{, }\AttributeTok{ylab =} \StringTok{"PIP"}\NormalTok{, }\AttributeTok{main =} \StringTok{"Variable Importance"}\NormalTok{)}
\FunctionTok{plot}\NormalTok{(variable\_selection\_2}\SpecialCharTok{$}\NormalTok{varimp, }
     \AttributeTok{col =} \FunctionTok{ifelse}\NormalTok{(}\DecValTok{1}\SpecialCharTok{:}\DecValTok{250} \SpecialCharTok{\textless{}} \DecValTok{6}\NormalTok{, }\StringTok{"\#386CB0"}\NormalTok{, }\StringTok{"\#7FC97F"}\NormalTok{), }\AttributeTok{pch =} \DecValTok{20}\NormalTok{,}
     \AttributeTok{xlab =} \StringTok{"Predictor"}\NormalTok{, }\AttributeTok{ylab =} \StringTok{"PIP"}\NormalTok{, }\AttributeTok{main =} \StringTok{"Variable Importance"}\NormalTok{)}
\end{Highlighting}
\end{Shaded}

We see that placing a prior on \(s\) leads to a much sharper transition
between the relevant and irrelevant predictors in terms of variable
importance. When the data generating mechanism is sparse this leads to
better predictions, as the ensemble can dedicate more splitting rules to
the relevant predictors; for example, we see in Figure
\ref{fig:plotvarimp} that placing a prior on \(s\) also allows the
ensemble to use more splitting rules to capture the interaction between
\(x_1\) and \(x_2\).

\hypertarget{embedding-softbart-into-other-models}{%
\section{Embedding SoftBart Into Other
Models}\label{embedding-softbart-into-other-models}}

Beyond allowing for the use of soft decision trees, \texttt{SoftBart}
has the unique advantage of making it easy for researchers to embed a
\texttt{SoftBart} model into other models within \texttt{R} without
having to modify the underlying \texttt{C++} code. This allows users to
seamlessly extend \texttt{SoftBart} to any setting in which the Bayesian
backfitting algorithm can be applied. Additionally, \texttt{SoftBart}
allows users to construct \emph{multiple} forests, which is required to
implement (for example) the \emph{Bayesian causal forest} (BCF) model of
\citet{hahn2020bayesian}.

Users can construct a decision tree ensemble using the function
\texttt{MakeForest()}, which returns a pointer to an \texttt{Rcpp}
\citep{eddelbuettel2018extending} data structure called an
\texttt{Rcpp\_Forest}. \texttt{Rcpp\_Forest}s can be interacted with
using the \texttt{\$} operator, and a full list of available functions
is given in the documentation for \texttt{MakeForest()}. Some important
functions include:

\begin{itemize}
\item
  \texttt{forest\$do\_gibbs(X,\ Y,\ X\_test,\ i)} runs \(i\) iterations
  of the Bayesian backfitting algorithm with covariate matrix \texttt{X}
  and outcome \texttt{Y}. It returns predictions on the test set of
  covariates \texttt{X\_test}. It also changes the data structure
  itself, with the state of \texttt{forest} now reflecting the forest
  after an additional \(i\) iterations have been run. The related
  function \texttt{do\_gibbs\_weights(X,\ Y,\ weights,\ X\_test)} runs a
  heteroskedastic version of the Bayesian backfitting update, with the
  error variance for each observation proportional to
  \texttt{1/weights}.
\item
  \texttt{forest\$do\_predict(X)} returns predictions for the covariate
  matrix \texttt{X} at the \emph{current state} of \texttt{forest}.
\item
  \texttt{forest\$get\_sigma()} and \texttt{forest\$set\_sigma()} allow
  us to retrieve and change the error variance \(\sigma^2\) assumed in
  the regression model of \texttt{Y} on \texttt{X}. This can be useful
  if we have multiple forests being updated with only one error variance
  parameter.
\end{itemize}

We give some examples below. For convenience, more developed versions of
the basic functions we write are included in \texttt{SoftBart} as the
\texttt{softbart\_probit()}, \texttt{vc\_softbart\_regression()}, and
\texttt{goftbartbart\_regression()} functions.

\hypertarget{probit-regression-with-data-augmentation}{%
\subsection{Probit Regression with Data
Augmentation}\label{probit-regression-with-data-augmentation}}

It is straight-forward using an \texttt{Rcpp\_Forest} object to
implement data augmentation algorithms, such as the algorithm of
\citet{albert1993bayesian}. The nonparametric probit regression model
takes
\([Y_i \mid X_i = x] \sim \operatorname{Bernoulli}[\Phi\{r(x)\}]\), and
this model can be expressed in terms of latent variables as \[
  Y_i = I(Z_i > 0) \qquad \text{where} \qquad Z_i \sim \operatorname{Normal}\{r(X_i), 1\}.
\] Let \(\operatorname{Normal}(\mu, \sigma^2, A)\) denote the normal
distribution truncated to the set \(A\) and let \(A_0 = (-\infty, 0)\)
and \(A_1 = (0, \infty)\). The data augmentation algorithm of
\citet{albert1993bayesian} alternates between (i) sampling the
unobserved latent variables
\(Z_i \sim \operatorname{Normal}\{r(X_i), 1, A_{Y_i}\}\) and (ii)
updating \(r(X_i)\) via Bayesian backfitting with the \(Z_i\)'s as the
outcomes. The first step can be accomplished in \texttt{R} using the
\texttt{rtruncnorm()} function in the \texttt{truncnorm} package
\citep{mersmann2018truncnorm}.

In implementing this algorithm, it is important to both set
\(\sigma = 1\) in our \texttt{Rcpp\_Forest} and to ensure that
\(\sigma\) is not updated, since the variance of \(Z_i\) is fixed at
\(1\). The following basic function will fit the probit model using
appropriate default values for the hyperparameters:

\begin{Shaded}
\begin{Highlighting}[]
\NormalTok{fit\_probit }\OtherTok{\textless{}{-}} \ControlFlowTok{function}\NormalTok{(X, Y, X\_test, num\_tree, num\_iter) \{}
  
  \DocumentationTok{\#\# Construct forest}
\NormalTok{  hypers }\OtherTok{\textless{}{-}} \FunctionTok{Hypers}\NormalTok{(X, Y, }\AttributeTok{k =} \DecValTok{1}\SpecialCharTok{/}\DecValTok{6}\NormalTok{, }
                   \AttributeTok{num\_tree =}\NormalTok{ num\_tree, }\AttributeTok{sigma\_hat =} \DecValTok{1}\NormalTok{)}
\NormalTok{  opts }\OtherTok{\textless{}{-}} \FunctionTok{Opts}\NormalTok{(}\AttributeTok{update\_sigma =} \ConstantTok{FALSE}\NormalTok{)}
  
\NormalTok{  probit\_forest }\OtherTok{\textless{}{-}} \FunctionTok{MakeForest}\NormalTok{(hypers, opts)}
  
  \DocumentationTok{\#\# Store the output}
\NormalTok{  r\_train }\OtherTok{\textless{}{-}} \FunctionTok{matrix}\NormalTok{(}\AttributeTok{nrow =}\NormalTok{ num\_iter, }\AttributeTok{ncol =} \FunctionTok{nrow}\NormalTok{(X))}
\NormalTok{  r\_test  }\OtherTok{\textless{}{-}} \FunctionTok{matrix}\NormalTok{(}\AttributeTok{nrow =}\NormalTok{ num\_iter, }\AttributeTok{ncol =}\NormalTok{ X\_test)}
  
  \DocumentationTok{\#\# Initialize chain}
\NormalTok{  r       }\OtherTok{\textless{}{-}}\NormalTok{ probit\_forest}\SpecialCharTok{$}\FunctionTok{do\_predict}\NormalTok{(X)}
\NormalTok{  upper   }\OtherTok{\textless{}{-}} \FunctionTok{ifelse}\NormalTok{(Y }\SpecialCharTok{==} \DecValTok{0}\NormalTok{, }\DecValTok{0}\NormalTok{, }\ConstantTok{Inf}\NormalTok{)}
\NormalTok{  lower   }\OtherTok{\textless{}{-}} \FunctionTok{ifelse}\NormalTok{(Y }\SpecialCharTok{==} \DecValTok{0}\NormalTok{, }\SpecialCharTok{{-}}\ConstantTok{Inf}\NormalTok{, }\DecValTok{0}\NormalTok{)}
\NormalTok{  Z     }\OtherTok{\textless{}{-}}\NormalTok{ truncnorm}\SpecialCharTok{::}\FunctionTok{rtruncnorm}\NormalTok{(}\AttributeTok{n =} \FunctionTok{length}\NormalTok{(Y\_probit\_train), }
                               \AttributeTok{a =}\NormalTok{ lower, }\AttributeTok{b =}\NormalTok{ upper, }\AttributeTok{mean =}\NormalTok{ r, }\AttributeTok{sd =} \DecValTok{1}\NormalTok{)}
  
  \DocumentationTok{\#\# Do MCMC}
  \ControlFlowTok{for}\NormalTok{(i }\ControlFlowTok{in} \DecValTok{1}\SpecialCharTok{:}\NormalTok{num\_iter) \{}
\NormalTok{    r }\OtherTok{\textless{}{-}}\NormalTok{ probit\_forest}\SpecialCharTok{$}\FunctionTok{do\_gibbs}\NormalTok{(X, Z, X, }\DecValTok{1}\NormalTok{)}
\NormalTok{    Z }\OtherTok{\textless{}{-}}\NormalTok{ truncnorm}\SpecialCharTok{::}\FunctionTok{rtruncnorm}\NormalTok{(}\AttributeTok{n =} \FunctionTok{length}\NormalTok{(Y\_probit\_train), }
                               \AttributeTok{a =}\NormalTok{ lower, }\AttributeTok{b =}\NormalTok{ upper, }\AttributeTok{mean =}\NormalTok{ r, }\AttributeTok{sd =} \DecValTok{1}\NormalTok{)}
\NormalTok{    r\_train[i,] }\OtherTok{\textless{}{-}}\NormalTok{ r}
\NormalTok{    r\_test[i,] }\OtherTok{\textless{}{-}}\NormalTok{ probit\_forest}\SpecialCharTok{$}\FunctionTok{do\_predict}\NormalTok{(X\_test)}
\NormalTok{  \}}
  
  \DocumentationTok{\#\# Return results}
  \FunctionTok{return}\NormalTok{(}\FunctionTok{list}\NormalTok{(}\AttributeTok{r\_train =}\NormalTok{ r\_train, }\AttributeTok{r\_test =}\NormalTok{ r\_test))}
\NormalTok{\}}
\end{Highlighting}
\end{Shaded}

The corresponding function \texttt{softbart\_probit()} in
\texttt{SoftBart} is more complicated, but ultimately uses the same
implementation as the above code.

I now test this function using data from the probit regression model
with \(r(x) = \frac{3}{5}\{r_0(x) - 14\}\) where \(r_0(x)\) is the same
function used in our semiparametric regression illustration. I first
generate the data:

\begin{Shaded}
\begin{Highlighting}[]
\FunctionTok{set.seed}\NormalTok{(}\DecValTok{77887}\NormalTok{)}
\NormalTok{r\_probit\_train }\OtherTok{\textless{}{-}} \DecValTok{3}\SpecialCharTok{*}\NormalTok{(training\_data}\SpecialCharTok{$}\NormalTok{mu }\SpecialCharTok{{-}} \DecValTok{14}\NormalTok{) }\SpecialCharTok{/} \DecValTok{5}
\NormalTok{r\_probit\_test  }\OtherTok{\textless{}{-}} \DecValTok{3}\SpecialCharTok{*}\NormalTok{(test\_data}\SpecialCharTok{$}\NormalTok{mu }\SpecialCharTok{{-}} \DecValTok{14}\NormalTok{) }\SpecialCharTok{/} \DecValTok{5}
\NormalTok{p\_train        }\OtherTok{\textless{}{-}} \FunctionTok{pnorm}\NormalTok{(r\_probit\_train)}
\NormalTok{p\_test        }\OtherTok{\textless{}{-}} \FunctionTok{pnorm}\NormalTok{(r\_probit\_test)}
\NormalTok{Y\_probit\_train }\OtherTok{\textless{}{-}} \FunctionTok{rbinom}\NormalTok{(}\FunctionTok{length}\NormalTok{(p\_train), }\AttributeTok{size =} \DecValTok{1}\NormalTok{, }\AttributeTok{prob =}\NormalTok{ p\_train)}
\NormalTok{Y\_probit\_test }\OtherTok{\textless{}{-}} \FunctionTok{rbinom}\NormalTok{(}\FunctionTok{length}\NormalTok{(p\_test), }\AttributeTok{size =} \DecValTok{1}\NormalTok{, }\AttributeTok{prob =}\NormalTok{ p\_test)}
\end{Highlighting}
\end{Shaded}

I then fit the model:

\begin{Shaded}
\begin{Highlighting}[]
\FunctionTok{set.seed}\NormalTok{(}\DecValTok{1903}\NormalTok{)}
\NormalTok{fitted\_probit }\OtherTok{\textless{}{-}} \FunctionTok{fit\_probit}\NormalTok{(}\AttributeTok{X =}\NormalTok{ X\_train, }\AttributeTok{Y =}\NormalTok{ Y\_probit\_train,}
                            \AttributeTok{X\_test =}\NormalTok{ X\_test, }
                            \AttributeTok{num\_tree =} \DecValTok{20}\NormalTok{, }\AttributeTok{num\_iter =} \DecValTok{5000}\NormalTok{)}
\end{Highlighting}
\end{Shaded}

\begin{figure}[t]

{\centering \includegraphics{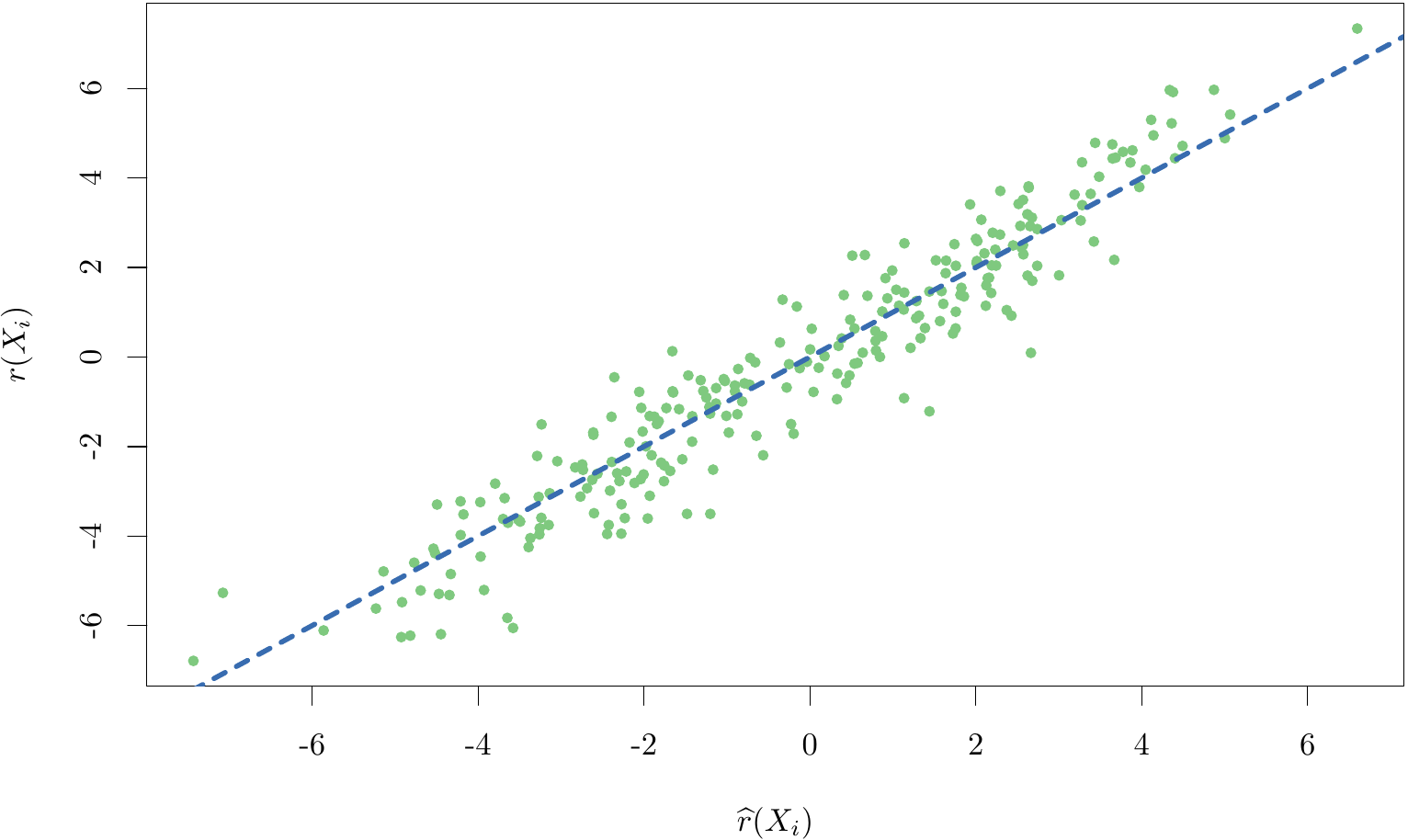} 

}

\caption{Results from fitting the probit regression model to the simulated data.}\label{fig:tikzy}
\end{figure}

The following code plots the estimated values of \(r(X_i)\) from the
\texttt{fit\_probit()} output to against their true values, with the
results in Figure \ref{fig:tikzy}:

\begin{Shaded}
\begin{Highlighting}[]
\FunctionTok{plot}\NormalTok{(}\FunctionTok{colMeans}\NormalTok{(fitted\_probit}\SpecialCharTok{$}\NormalTok{r\_train), r\_probit\_train, }
     \AttributeTok{xlab =} \StringTok{"$}\SpecialCharTok{\textbackslash{}\textbackslash{}}\StringTok{widehat r(X\_i)$"}\NormalTok{, }\AttributeTok{ylab =} \StringTok{"$r(X\_i)$"}\NormalTok{, }
     \AttributeTok{pch =} \DecValTok{20}\NormalTok{, }\AttributeTok{col =} \StringTok{"\#7FC97F"}\NormalTok{)}
\FunctionTok{abline}\NormalTok{(}\AttributeTok{a =} \DecValTok{0}\NormalTok{, }\AttributeTok{b =} \DecValTok{1}\NormalTok{, }\AttributeTok{col =} \StringTok{"\#386CB0"}\NormalTok{, }\AttributeTok{lwd =} \DecValTok{4}\NormalTok{, }\AttributeTok{lty =} \DecValTok{2}\NormalTok{)}
\end{Highlighting}
\end{Shaded}

The same functionality is available in the package with the
\texttt{softbart\_probit()} function, which can be fit as follows:

\begin{Shaded}
\begin{Highlighting}[]
\NormalTok{probit\_data }\OtherTok{\textless{}{-}} \FunctionTok{data.frame}\NormalTok{(}\AttributeTok{X =}\NormalTok{ X\_train, }
                          \AttributeTok{Y =} \FunctionTok{factor}\NormalTok{(Y\_probit\_train, }\AttributeTok{levels =} \FunctionTok{c}\NormalTok{(}\DecValTok{0}\NormalTok{,}\DecValTok{1}\NormalTok{)))}
\NormalTok{probit\_test }\OtherTok{\textless{}{-}} \FunctionTok{data.frame}\NormalTok{(}\AttributeTok{X =}\NormalTok{ X\_test, }
                          \AttributeTok{Y =} \FunctionTok{factor}\NormalTok{(Y\_probit\_test, }\AttributeTok{levels =} \FunctionTok{c}\NormalTok{(}\DecValTok{0}\NormalTok{,}\DecValTok{1}\NormalTok{)))}

\NormalTok{fitted\_probit }\OtherTok{\textless{}{-}} \FunctionTok{softbart\_probit}\NormalTok{(Y }\SpecialCharTok{\textasciitilde{}}\NormalTok{ ., }\AttributeTok{data =}\NormalTok{ probit\_data, }
                                 \AttributeTok{test\_data =}\NormalTok{ probit\_test, }\AttributeTok{verbose =} \ConstantTok{FALSE}\NormalTok{)}
\end{Highlighting}
\end{Shaded}

\hypertarget{a-varying-coefficient-bart-model-and-a-bayesian-causal-forest}{%
\subsection{A Varying Coefficient BART Model and a Bayesian Causal
Forest}\label{a-varying-coefficient-bart-model-and-a-bayesian-causal-forest}}

The \emph{varying coefficient BART} (VC-BART) model of
\citet{deshpande2020vcbart} assumes a linear relationship in a covariate
of interest \(Z_i\), with the regression coefficient possibly depending
on the other covariates: \begin{align}
  \label{eq:vcbart}
  Y_i = \alpha(X_i) + Z_i \, \beta(X_i) + \epsilon_i.
\end{align} Here, \(Z_i\) is the covariate of interest, \(X_i\) is a
vector of other covariates, and
\(\epsilon_i \sim \operatorname{Normal}(0, \sigma^2)\). This model can
be fit via a two-stage Gibbs sampler by sampling from the distributions
of \([\alpha, \sigma^2 \mid \beta, \text{Data}]\) and
\([\beta \mid \alpha, \sigma^2, \text{Data}]\). We can derive the update
for \((\alpha, \sigma^2)\) by forming the residuals
\(R_{\alpha i} = Y_i - Z_i \, \beta(X_i) = \alpha(X_i) + \epsilon_i\),
which follow the usual BART model. An update for \(\beta(\cdot)\) can be
derived similarly by noting that \begin{align*}
  \frac{Y_i - \alpha(X_i)}{Z_i} 
  \equiv
  R_{\beta i}
  \sim \operatorname{Normal}\left\{\beta(X_i), \frac{\sigma^2}{Z_i^2}\right\}.
\end{align*} Hence \(R_{\beta i}\) follows a \emph{heteroskedastic BART}
model, with weights \(w_i = Z_i^2\). \texttt{Rcpp\_Forest} objects allow
users to specify a vector of weights using the
\texttt{\$do\_gibbs\_weighted()} method, and can therefore handle the
update for \(\beta\) as well. A short function for fitting the VC-BART
model is given in Appendix \ref{code-for-the-vc-bart-model}; the main
component of this code is the pair of updates for \((\alpha, \sigma^2)\)
and \(\beta\):

\begin{Shaded}
\begin{Highlighting}[]
\DocumentationTok{\#\# Update alpha}
\NormalTok{R\_alpha }\OtherTok{\textless{}{-}}\NormalTok{ Y }\SpecialCharTok{{-}}\NormalTok{ Z }\SpecialCharTok{*}\NormalTok{ beta}
\NormalTok{alpha }\OtherTok{\textless{}{-}}\NormalTok{ alpha\_forest}\SpecialCharTok{$}\FunctionTok{do\_gibbs}\NormalTok{(X, R\_alpha, X, }\DecValTok{1}\NormalTok{)}
\NormalTok{sigma }\OtherTok{\textless{}{-}}\NormalTok{ alpha\_forest}\SpecialCharTok{$}\FunctionTok{get\_sigma}\NormalTok{()}

\DocumentationTok{\#\# Update beta}
\NormalTok{R\_beta }\OtherTok{\textless{}{-}}\NormalTok{ (Y  }\SpecialCharTok{{-}}\NormalTok{ alpha) }\SpecialCharTok{/}\NormalTok{ Z}
\NormalTok{beta\_forest}\SpecialCharTok{$}\FunctionTok{set\_sigma}\NormalTok{(sigma)}
\NormalTok{beta }\OtherTok{\textless{}{-}}\NormalTok{ beta\_forest}\SpecialCharTok{$}\FunctionTok{do\_gibbs\_weight}\NormalTok{(X, R\_beta, Z}\SpecialCharTok{\^{}}\DecValTok{2}\NormalTok{, X, }\DecValTok{1}\NormalTok{)}
\end{Highlighting}
\end{Shaded}

Note that it is important that \texttt{beta\_forest} and
\texttt{alpha\_forest} share the same value of \(\sigma\) internally,
hence the call of \texttt{beta\_forest\$set\_sigma(sigma)}. This design
pattern is common in models where multiple forests are used.

\begin{figure}[t]

{\centering \includegraphics[width=.9\textwidth,]{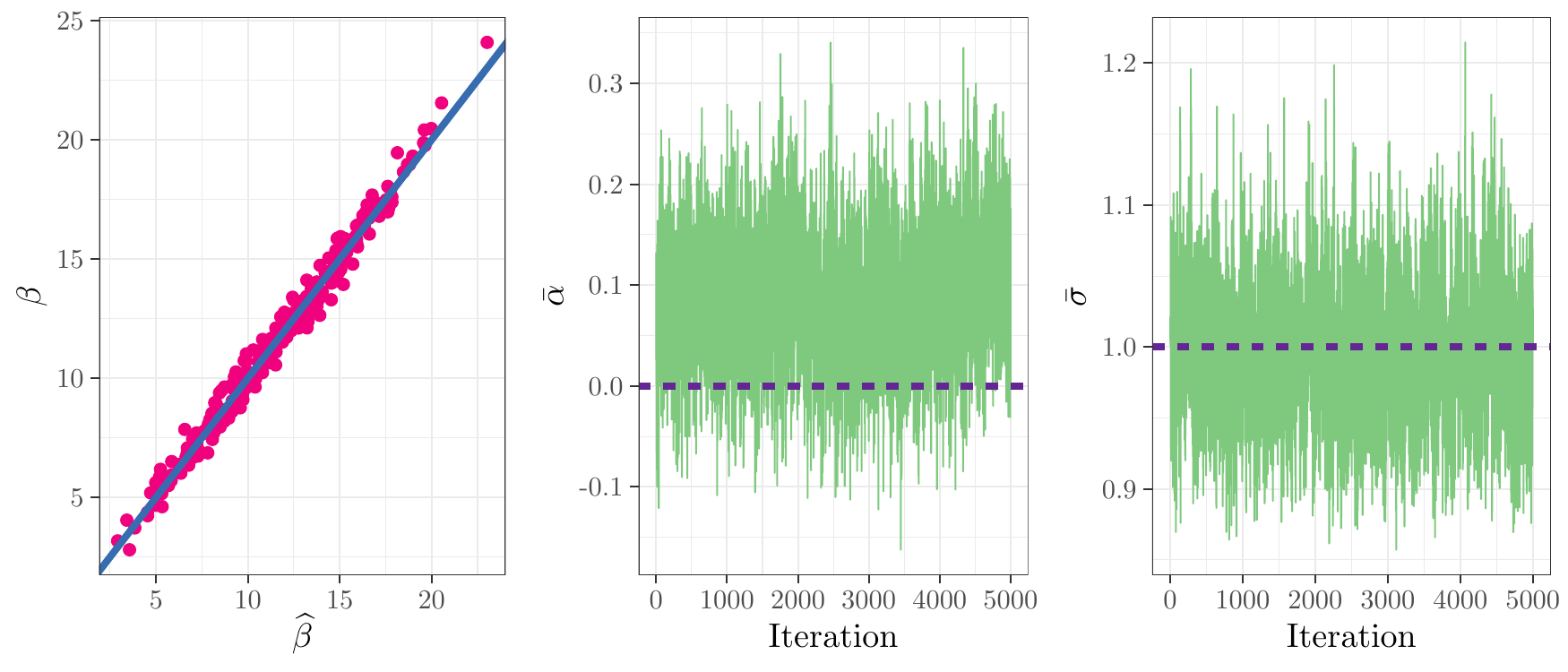} 

}

\caption{Fit of the varying coefficient model to the simulated data. Left: plot
of $\widehat \beta(X_i)$ against $\beta(X_i)$ for each observation. Middle:
traceplot of $\bar \alpha$. Right: traceplot of $\sigma$.}\label{fig:vc-results}
\end{figure}

To evaluate the model, we generate data that takes \(\beta(x) = r(x)\)
and \(\alpha(x) \equiv 0\), where \(r(x)\) is given in
\eqref{eq:friedman}. Figure \ref{fig:vc-results} shows that our VC-BART
model effectively estimates \(\beta(X_i)\), \(\sigma^2\), and sets
\(\bar \alpha = \frac{1}{N} \sum_i \alpha(X_i) \approx 0\).

The varying coefficient model contains the \emph{Bayesian causal forest}
(BCF) model of \citet{hahn2020bayesian} as a special case. This model
takes the outcome to be the \emph{observed outcome}
\(Y_i \equiv Y_i(A_i)\) of a pair of \emph{potential outcomes}
\(\{Y_i(0), Y_i(1)\}\) under a binary treatment variable
\(a \in \{0, 1\}\). A BCF specifies \[
  Y_i(a) = \mu(X_i) + a \, \tau(X_i) + \epsilon_i, \qquad \epsilon_i \sim \operatorname{Normal}(0, \sigma^2).
\] When applying BCFs, one is typically interested in both the
\emph{population average causal effect} (PACE), given by
\(\tau = \mathbb E\{Y_i(1) - Y_i(0)\}\) or the \emph{conditional average
causal effect} (CACE) given by
\(\tau(x) = \mathbb E\{Y_i(1) - Y_i(0) \mid X_i = x\}\). A simple way to
implement a BCF is to define \(Z_i = \nicefrac{1}{2} - A_i\) and fit the
VC-BART model \eqref{eq:vcbart}. Under the common \emph{ignorability}
assumption that the treatment \(Z_i\) is independent of the potential
outcomes \(\{Y_i(0), Y_i(1)\}\) conditional on the covariates \(X_i\),
it is then easy to show that \(\beta(x) = \tau(x)\) for this choice of
\(Z_i\).

\hypertarget{the-general-bart-model}{%
\subsection{The General BART Model}\label{the-general-bart-model}}

Finally, I show how \texttt{SoftBart} can be used to implement the
\emph{general BART model} described by \citet{tan2019bayesian}. This is
effectively a partial linear model \[
  Y_i = r(X_i) + Z_i^\top \beta + \epsilon_i,
\] which is straight-forward to also extend to probit outcomes. This
model can also be used to encode a \emph{mixed effects model} when
\(\beta\) is a vector of random effects; for simplicity, I will take
\(\beta\) to be a set of fixed effects, with a flat prior on \(\beta\).

Code for fitting this model is given in Appendix
\ref{code-for-the-general-bart-model}, with the relevant updates being
given by the lines

\begin{figure}[t]

{\centering \includegraphics[width=0.9\textwidth,]{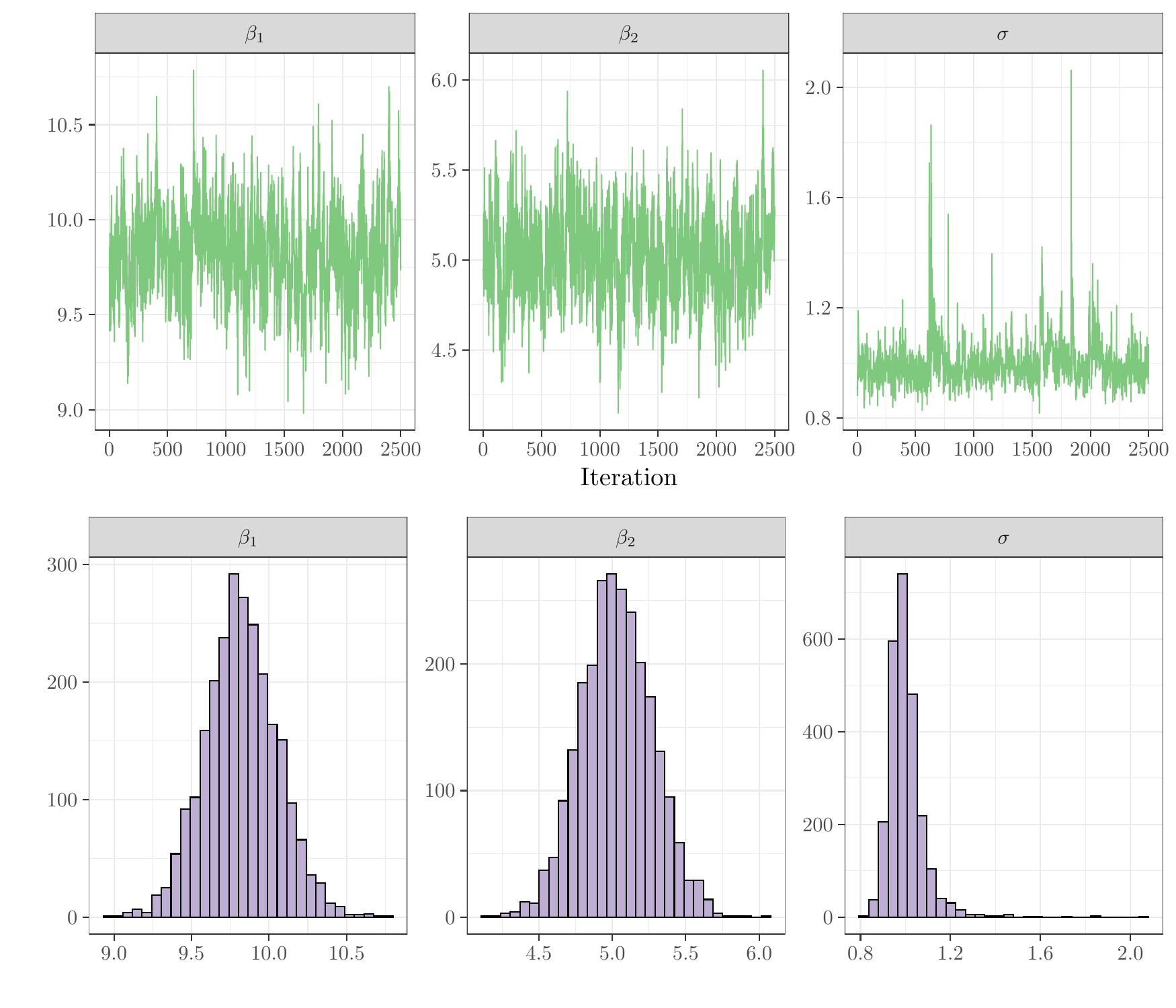} 

}

\caption{Traceplots (top) and posterior histograms (bottom) associated with the general
BART model fit to the simulation setting of \eqref{eq:friedman}.}\label{fig:gbartfig}
\end{figure}

\begin{Shaded}
\begin{Highlighting}[]
\DocumentationTok{\#\# Update beta}
\NormalTok{R }\OtherTok{\textless{}{-}}\NormalTok{ Y\_train }\SpecialCharTok{{-}}\NormalTok{ r\_train}
\NormalTok{beta }\OtherTok{\textless{}{-}} \FunctionTok{update\_beta}\NormalTok{(R, Z\_train, sigma}\SpecialCharTok{\^{}}\DecValTok{2}\NormalTok{)}
    
\DocumentationTok{\#\# Update forest and sigma}
\NormalTok{R }\OtherTok{\textless{}{-}}\NormalTok{ Y\_train }\SpecialCharTok{{-}} \FunctionTok{as.numeric}\NormalTok{(Z\_train }\SpecialCharTok{\%*\%}\NormalTok{ beta)}
\NormalTok{r\_train }\OtherTok{\textless{}{-}}\NormalTok{ forest}\SpecialCharTok{$}\FunctionTok{do\_gibbs}\NormalTok{(X\_train, R, X\_train, }\DecValTok{1}\NormalTok{) }\SpecialCharTok{\%\textgreater{}\%} \FunctionTok{as.numeric}\NormalTok{()}
\NormalTok{sigma }\OtherTok{\textless{}{-}}\NormalTok{ forest}\SpecialCharTok{$}\FunctionTok{get\_sigma}\NormalTok{()}
\end{Highlighting}
\end{Shaded}

where the line
\texttt{beta\ \textless{}-\ update\_beta(R,\ Z\_train,\ sigma\^{}2)}
samples \(\beta\) from its full conditional \[
  \beta \sim \operatorname{Normal}\{(\bm Z^\top \bm Z)^{-1} \bm Z^\top \bm R, \sigma^2 (\bm Z^\top \bm Z)^{-1}\}.
\] I illustrate the use of this code by fitting data generated from the
semiparametric Gaussian regression model under \eqref{eq:friedman}, but
now taking into account the fact that \(X_4\) and \(X_5\) have linear
effects. Traceplots and posterior histograms for the parameters
\((\beta_1, \beta_2, \sigma)\) (with ground truth values \((10, 5, 1)\))
are given in Figure \ref{fig:gbartfig}, where \(\beta_1\) is the
regression coefficient for \(X_4\) and \(\beta_2\) is the regression
coefficient for \(X_5\). We see that the partial linear model is capable
of estimating both of the regression coefficients and the error variance
accurately and that the chain mixes well.

\hypertarget{illustration-diamonds}{%
\section{Illustration: Diamonds}\label{illustration-diamonds}}

I illustrate the use of \texttt{SoftBart} on the \texttt{diamonds}
dataset available in the \texttt{IIS} package \citep{schneider2017IIS}:

\begin{Shaded}
\begin{Highlighting}[]
\FunctionTok{library}\NormalTok{(IIS)}
\FunctionTok{data}\NormalTok{(}\StringTok{"diamonds\_carats\_color\_cost"}\NormalTok{)}
\NormalTok{diamonds\_pre }\OtherTok{\textless{}{-}}\NormalTok{ diamonds\_carats\_color\_cost}
\FunctionTok{head}\NormalTok{(diamonds\_pre)}
\end{Highlighting}
\end{Shaded}

\begin{verbatim}
##   carat color clarity certification_body price
## 1   0.3     D     VS2                GIA  1302
## 2   0.3     E     VS1                GIA  1510
## 3   0.3     G    VVS1                GIA  1510
## 4   0.3     G     VS1                GIA  1260
## 5  0.31     D     VS1                GIA  1641
## 6  0.31     E     VS1                GIA  1555
\end{verbatim}

\citet{linero2017abayesian} showed that \texttt{SoftBart} performs
better than competing methods (random forests, gradient boosted trees,
BART, and the lasso) on this dataset. I fit a semiparametric regression
model using \texttt{log(price)} as the outcome and with the remaining
variables used as predictors.

\begin{Shaded}
\begin{Highlighting}[]
\FunctionTok{set.seed}\NormalTok{(}\DecValTok{77777}\NormalTok{)}
\NormalTok{diamonds }\OtherTok{\textless{}{-}}\NormalTok{ diamonds\_pre }\SpecialCharTok{\%\textgreater{}\%} 
  \FunctionTok{mutate}\NormalTok{(}\AttributeTok{logprice =} \FunctionTok{log}\NormalTok{(}\FunctionTok{as.numeric}\NormalTok{(}\FunctionTok{as.character}\NormalTok{(price))), }
         \AttributeTok{carat =} \FunctionTok{as.numeric}\NormalTok{(}\FunctionTok{as.character}\NormalTok{(carat))) }\SpecialCharTok{\%\textgreater{}\%} \FunctionTok{select}\NormalTok{(}\SpecialCharTok{{-}}\NormalTok{price)}

\NormalTok{opts }\OtherTok{\textless{}{-}} \FunctionTok{Opts}\NormalTok{(}\AttributeTok{num\_burn =} \DecValTok{5000}\NormalTok{, }\AttributeTok{num\_save =} \DecValTok{2500}\NormalTok{, }\AttributeTok{num\_thin =} \DecValTok{4}\NormalTok{)}
\NormalTok{fitted\_diamonds }\OtherTok{\textless{}{-}} \FunctionTok{softbart\_regression}\NormalTok{(logprice }\SpecialCharTok{\textasciitilde{}}\NormalTok{ ., }\AttributeTok{data =}\NormalTok{ diamonds, }
                                       \AttributeTok{test\_data =}\NormalTok{ diamonds,}
                                       \AttributeTok{opts =}\NormalTok{ opts)}
\end{Highlighting}
\end{Shaded}

This model has four predictors: \texttt{carat}, \texttt{clarity},
\texttt{color}, and \texttt{certification\_body}. Examining the
posterior inclusion probabilities, all of the variables are included in
the median probability model with the exception of
\texttt{certification\_body}:

\begin{Shaded}
\begin{Highlighting}[]
\FunctionTok{posterior\_probs}\NormalTok{(fitted\_diamonds)[[}\StringTok{"post\_probs"}\NormalTok{]]}
\end{Highlighting}
\end{Shaded}

\begin{verbatim}
##              carat              color            clarity certification_body 
##             1.0000             1.0000             1.0000             0.3704
\end{verbatim}

Next, I use the \texttt{partial\_dependence\_regression()} function to
visualize the partial dependence functions for the variables
\texttt{carat} and \texttt{clarity}. As \texttt{carat} measures the
weight of a diamond, we expect that \(\operatorname{PD}_{\text{carat}}\)
should be increasing in \texttt{carat}, as larger diamonds should, all
other things being equal, be more expensive:

\begin{Shaded}
\begin{Highlighting}[]
\NormalTok{pd\_clarity }\OtherTok{\textless{}{-}} \FunctionTok{partial\_dependence\_regression}\NormalTok{(}
  \AttributeTok{fit =}\NormalTok{ fitted\_diamonds, }
  \AttributeTok{test\_data =}\NormalTok{ diamonds, }
  \AttributeTok{var\_str =} \StringTok{"clarity"}\NormalTok{, }
  \AttributeTok{grid =} \FunctionTok{unique}\NormalTok{(diamonds}\SpecialCharTok{$}\NormalTok{clarity)}
\NormalTok{)}

\NormalTok{pd\_carat }\OtherTok{\textless{}{-}} \FunctionTok{partial\_dependence\_regression}\NormalTok{(}
  \AttributeTok{fit =}\NormalTok{ fitted\_diamonds, }
  \AttributeTok{test\_data =}\NormalTok{ diamonds, }
  \AttributeTok{var\_str =} \StringTok{"carat"}\NormalTok{, }
  \AttributeTok{grid =} \FunctionTok{unique}\NormalTok{(diamonds}\SpecialCharTok{$}\NormalTok{carat)}
\NormalTok{)}
\end{Highlighting}
\end{Shaded}

\begin{figure}[t]

{\centering \includegraphics[width=1\textwidth,]{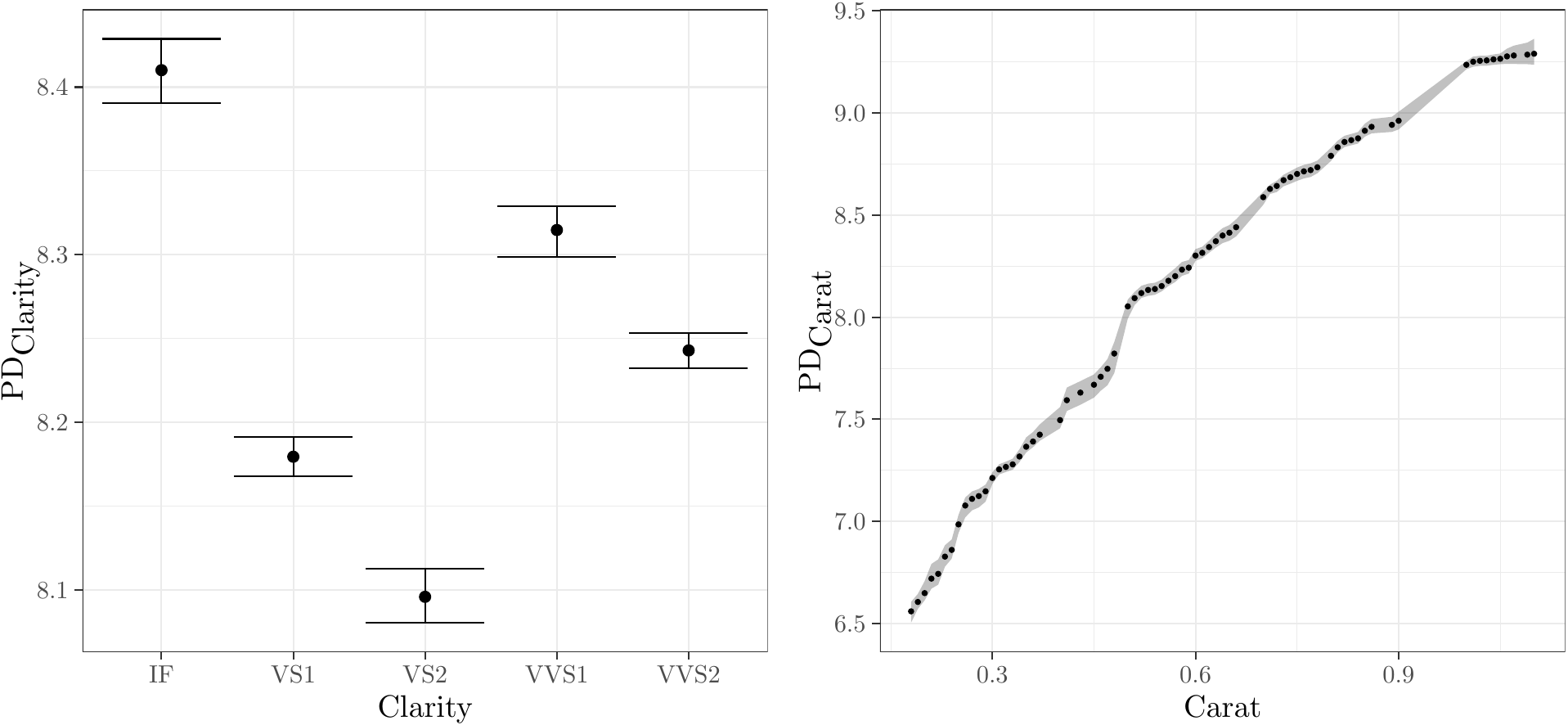} 

}

\caption{Estimates and credible intervals for the posterior dependence function of Clarity (left) and Carat(right).}\label{fig:pdpdiamond}
\end{figure}

The code below plots the posterior mean and 95\% credible bands for the
partial dependence functions (see Figure \ref{fig:pdpdiamond}):

\begin{Shaded}
\begin{Highlighting}[]
\NormalTok{LCL }\OtherTok{\textless{}{-}} \ControlFlowTok{function}\NormalTok{(x) }\FunctionTok{quantile}\NormalTok{(x, }\FloatTok{0.025}\NormalTok{)}
\NormalTok{UCL }\OtherTok{\textless{}{-}} \ControlFlowTok{function}\NormalTok{(x) }\FunctionTok{quantile}\NormalTok{(x, }\FloatTok{0.975}\NormalTok{)}

\NormalTok{pdp\_clarity }\OtherTok{\textless{}{-}} \FunctionTok{ggplot}\NormalTok{(pd\_clarity}\SpecialCharTok{$}\NormalTok{pred\_df, }\FunctionTok{aes}\NormalTok{(}\AttributeTok{x =}\NormalTok{ clarity, }\AttributeTok{y =}\NormalTok{ mu)) }\SpecialCharTok{+} 
  \FunctionTok{geom\_point}\NormalTok{(}\AttributeTok{stat =} \StringTok{"summary"}\NormalTok{, }\AttributeTok{fun =}\NormalTok{ mean) }\SpecialCharTok{+} 
  \FunctionTok{geom\_errorbar}\NormalTok{(}\AttributeTok{stat =} \StringTok{"summary"}\NormalTok{, }\AttributeTok{fun.min =}\NormalTok{ LCL, }\AttributeTok{fun.max =}\NormalTok{ UCL) }\SpecialCharTok{+} 
  \FunctionTok{xlab}\NormalTok{(}\StringTok{"Clarity"}\NormalTok{) }\SpecialCharTok{+} 
  \FunctionTok{ylab}\NormalTok{(}\StringTok{"$}\SpecialCharTok{\textbackslash{}\textbackslash{}}\StringTok{mbox\{PD\}\_\{}\SpecialCharTok{\textbackslash{}\textbackslash{}}\StringTok{mbox\{Clarity\}\}$"}\NormalTok{) }\SpecialCharTok{+} 
  \FunctionTok{theme\_bw}\NormalTok{()}

\NormalTok{pdp\_carat }\OtherTok{\textless{}{-}} \FunctionTok{ggplot}\NormalTok{(pd\_carat}\SpecialCharTok{$}\NormalTok{pred\_df, }\FunctionTok{aes}\NormalTok{(}\AttributeTok{x =}\NormalTok{ carat, }\AttributeTok{y =}\NormalTok{ mu)) }\SpecialCharTok{+} 
  \FunctionTok{geom\_ribbon}\NormalTok{(}\AttributeTok{stat =} \StringTok{"summary"}\NormalTok{, }
              \AttributeTok{fun.min =}\NormalTok{ LCL, }
              \AttributeTok{fun.max =}\NormalTok{ UCL, }
              \AttributeTok{alpha =} \FloatTok{0.3}
\NormalTok{  ) }\SpecialCharTok{+} 
  \FunctionTok{geom\_point}\NormalTok{(}\AttributeTok{stat =} \StringTok{"summary"}\NormalTok{, }
             \AttributeTok{fun =}\NormalTok{ mean, }
             \AttributeTok{size =} \FloatTok{0.5}
\NormalTok{  ) }\SpecialCharTok{+} 
  \FunctionTok{xlab}\NormalTok{(}\StringTok{"Carat"}\NormalTok{) }\SpecialCharTok{+} 
  \FunctionTok{ylab}\NormalTok{(}\StringTok{"$}\SpecialCharTok{\textbackslash{}\textbackslash{}}\StringTok{mbox\{PD\}\_\{}\SpecialCharTok{\textbackslash{}\textbackslash{}}\StringTok{mbox\{Carat\}\}$"}\NormalTok{) }\SpecialCharTok{+} 
  \FunctionTok{theme\_bw}\NormalTok{()}

\NormalTok{gridExtra}\SpecialCharTok{::}\FunctionTok{grid.arrange}\NormalTok{(pdp\_clarity, pdp\_carat, }\AttributeTok{nrow =} \DecValTok{1}\NormalTok{)}
\end{Highlighting}
\end{Shaded}

We see that there is very little uncertainty in the relationship between
\texttt{carat} and \texttt{log(price)}, although there are some values
of \texttt{carat} where we do not have much data. There is also a clear
effect of \texttt{clarity} on \texttt{log(price)}; in reality, the
variables \texttt{color} and \texttt{clarity} are ordinal, with the
model correctly ranking the possible values of \texttt{clarity} as IF
(best), VVS1, VVS2, VS1, and VS2 (worst).

\texttt{SoftBart} is, to the best of my knowledge, the only package for
fitting BART models that allows for partial dependence plots to be
computed for categorical variables with three or more levels, and hence
it would not be convenient to compute the partial dependence function
for \texttt{clarity} using other packages.

\hypertarget{discussion}{%
\section{Discussion}\label{discussion}}

Moving forward, it is my plan to expand \texttt{SoftBart} to include
recent developments in BART methodology. In ongoing work, the
\texttt{MakeForest()} functionality has been used to extent BCFs to
mediation analysis \citep{imai2010general} and panel data
\citep{bell2015explaining} settings.

There are a couple of important settings where neither \texttt{SoftBart}
nor most other existing BART packages are applicable. Several recent
applications of BART have made critical use of \emph{targeted smoothing}
over a variable of interest, such as time
\citep{starling2020bart, li2020adaptive, linero2021bayesian}, and, while
it would be feasible to do so, the functionality to do this has not been
added to \texttt{SoftBart} at this point. Additionally, recent work of
\citet{murray2021log} has allowed for the extension of BART to loglinear
models, gamma regression models \citep{linero2018shared},
heteroskedastic regression models \citep{pratola2017heteroscedastic},
and Cox survival models \citep{linero2021bayesian}, but the Bayesian
backfitting algorithms used for these extensions cannot be used with
soft decision trees. Finally, the trees used are strictly univariate,
and \texttt{SoftBart} does not allow for the use of multivariate
decision trees as used by \citet{linero2018shared}.

\appendix

\hypertarget{code-for-the-vc-bart-model}{%
\section{Code for the VC-BART Model}\label{code-for-the-vc-bart-model}}

The following function for fitting the VC-BART model takes as input two
forests \texttt{alpha\_forest} and \texttt{beta\_forest} constructed
using the \texttt{MakeForest()} function, along with a design matrix
\texttt{X}, an outcome \texttt{y} (both of which are assumed to have
been appropriately scaled), and a covariate to be treated linearly
\texttt{Z}. \texttt{SoftBart} includes broader functionality for the
VC-BART model via the \texttt{vc\_softbart\_regression()} function.

\begin{Shaded}
\begin{Highlighting}[]
\NormalTok{fit\_vc\_bart }\OtherTok{\textless{}{-}} \ControlFlowTok{function}\NormalTok{(alpha\_forest, beta\_forest, y, X, Z, num\_iter) \{}
  
  \DocumentationTok{\#\# Variables to save}
\NormalTok{  alpha\_out }\OtherTok{\textless{}{-}} \FunctionTok{matrix}\NormalTok{(}\ConstantTok{NA}\NormalTok{, }\AttributeTok{nrow =}\NormalTok{ num\_iter, }\AttributeTok{ncol =} \FunctionTok{nrow}\NormalTok{(X))}
\NormalTok{  beta\_out  }\OtherTok{\textless{}{-}} \FunctionTok{matrix}\NormalTok{(}\ConstantTok{NA}\NormalTok{, }\AttributeTok{nrow =}\NormalTok{ num\_iter, }\AttributeTok{ncol =} \FunctionTok{nrow}\NormalTok{(X))}
\NormalTok{  sigma\_out }\OtherTok{\textless{}{-}} \FunctionTok{numeric}\NormalTok{(num\_save)}
  
  \DocumentationTok{\#\# Initializing alpha vector}
\NormalTok{  alpha }\OtherTok{\textless{}{-}}\NormalTok{ alpha\_forest}\SpecialCharTok{$}\FunctionTok{do\_predict}\NormalTok{(X)}
  
  \ControlFlowTok{for}\NormalTok{(i }\ControlFlowTok{in} \DecValTok{1}\SpecialCharTok{:}\NormalTok{num\_iter) \{}
\NormalTok{    R }\OtherTok{\textless{}{-}}\NormalTok{ (y }\SpecialCharTok{{-}}\NormalTok{ alpha) }\SpecialCharTok{/}\NormalTok{ Z}
\NormalTok{    beta }\OtherTok{\textless{}{-}}\NormalTok{ beta\_forest}\SpecialCharTok{$}\FunctionTok{do\_gibbs\_weighted}\NormalTok{(X, R, Z}\SpecialCharTok{\^{}}\DecValTok{2}\NormalTok{, X, }\DecValTok{1}\NormalTok{)}
\NormalTok{    sigma }\OtherTok{\textless{}{-}}\NormalTok{ beta\_forest}\SpecialCharTok{$}\FunctionTok{get\_sigma}\NormalTok{()}
\NormalTok{    alpha\_forest}\SpecialCharTok{$}\FunctionTok{set\_sigma}\NormalTok{(sigma)}
\NormalTok{    R }\OtherTok{\textless{}{-}}\NormalTok{ (y }\SpecialCharTok{{-}}\NormalTok{ Z }\SpecialCharTok{*}\NormalTok{ beta)}
\NormalTok{    alpha }\OtherTok{\textless{}{-}}\NormalTok{ alpha\_forest}\SpecialCharTok{$}\FunctionTok{do\_gibbs}\NormalTok{(X, R, X, }\DecValTok{1}\NormalTok{)}
    
\NormalTok{    alpha\_out[i,] }\OtherTok{\textless{}{-}}\NormalTok{ alpha}
\NormalTok{    beta\_out[i,]  }\OtherTok{\textless{}{-}}\NormalTok{ beta}
\NormalTok{    sigma\_out[i]  }\OtherTok{\textless{}{-}}\NormalTok{ sigma}
\NormalTok{  \}}
  
\NormalTok{  mu\_out }\OtherTok{\textless{}{-}}\NormalTok{ alpha\_out }\SpecialCharTok{+} \FunctionTok{t}\NormalTok{(Z }\SpecialCharTok{*} \FunctionTok{t}\NormalTok{(beta\_out))}
  \FunctionTok{return}\NormalTok{(}\FunctionTok{list}\NormalTok{(}\AttributeTok{alpha =}\NormalTok{ alpha\_out, }\AttributeTok{beta =}\NormalTok{ beta\_out, }
              \AttributeTok{sigma =}\NormalTok{ sigma\_out, }\AttributeTok{mu =}\NormalTok{ mu\_out))}
\NormalTok{\}}
\end{Highlighting}
\end{Shaded}

\hypertarget{code-for-the-general-bart-model}{%
\section{Code for the General BART
Model}\label{code-for-the-general-bart-model}}

The following function does a conjugate update for a parameter \(\beta\)
in a Bayesian linear regression model
\(R_i = Z_i^\top \beta + \epsilon_i\) under a flat prior for \(\beta\).

\begin{Shaded}
\begin{Highlighting}[]
\NormalTok{update\_beta }\OtherTok{\textless{}{-}} \ControlFlowTok{function}\NormalTok{(R, Z, sigma\_sq) \{}
\NormalTok{  ZtR }\OtherTok{\textless{}{-}} \FunctionTok{t}\NormalTok{(Z) }\SpecialCharTok{\%*\%}\NormalTok{ R}
\NormalTok{  ZtZi }\OtherTok{\textless{}{-}} \FunctionTok{solve}\NormalTok{(}\FunctionTok{t}\NormalTok{(Z) }\SpecialCharTok{\%*\%}\NormalTok{ Z)}
\NormalTok{  beta\_hat }\OtherTok{\textless{}{-}}\NormalTok{ ZtZi }\SpecialCharTok{\%*\%}\NormalTok{ ZtR}
\NormalTok{  Sigma }\OtherTok{\textless{}{-}}\NormalTok{ sigma\_sq }\SpecialCharTok{*}\NormalTok{ ZtZi}
\NormalTok{  beta }\OtherTok{\textless{}{-}}\NormalTok{ MASS}\SpecialCharTok{::}\FunctionTok{mvrnorm}\NormalTok{(}\AttributeTok{n =} \DecValTok{1}\NormalTok{, }\AttributeTok{mu =}\NormalTok{ beta\_hat, }\AttributeTok{Sigma =}\NormalTok{ Sigma) }\SpecialCharTok{\%\textgreater{}\%} 
    \FunctionTok{as.numeric}\NormalTok{()}
  \FunctionTok{return}\NormalTok{(beta)}
\NormalTok{\}}
\end{Highlighting}
\end{Shaded}

Using the function to update \(\beta\), along with the functionality
from \texttt{SoftBart}, we can fit the general BART model. The following
function for fitting the generalized BART model takes as input a forest
\texttt{r\_forest} constructed using the \texttt{MakeForest()} function,
along with a design matrix \texttt{X}, an outcome \texttt{y} (both of
which are assumed to be scaled) and a design matrix \texttt{Z} of
covariates to be treated linearly. This functionality is available in
the \texttt{gsoftbart\_regression()} function in \texttt{SoftBart}.

\begin{Shaded}
\begin{Highlighting}[]
\NormalTok{fit\_gbart }\OtherTok{\textless{}{-}} \ControlFlowTok{function}\NormalTok{(r\_forest, y, X, Z, num\_iter) \{}
  
  \DocumentationTok{\#\# Variables to save}
\NormalTok{  r\_out }\OtherTok{\textless{}{-}} \FunctionTok{matrix}\NormalTok{(}\ConstantTok{NA}\NormalTok{, }\AttributeTok{nrow =}\NormalTok{ num\_iter, }\AttributeTok{ncol =} \FunctionTok{nrow}\NormalTok{(X))}
\NormalTok{  beta\_out }\OtherTok{\textless{}{-}} \FunctionTok{matrix}\NormalTok{(}\ConstantTok{NA}\NormalTok{, }\AttributeTok{nrow =}\NormalTok{ num\_iter, }\AttributeTok{ncol =} \FunctionTok{ncol}\NormalTok{(Z))}
\NormalTok{  sigma\_out }\OtherTok{\textless{}{-}} \FunctionTok{numeric}\NormalTok{(num\_save)}
\NormalTok{  eta\_out }\OtherTok{\textless{}{-}} \FunctionTok{matrix}\NormalTok{(}\ConstantTok{NA}\NormalTok{, }\AttributeTok{nrow =}\NormalTok{ num\_iter, }\AttributeTok{ncol =} \FunctionTok{nrow}\NormalTok{(X))}

  \DocumentationTok{\#\# Initializing}
\NormalTok{  r }\OtherTok{\textless{}{-}}\NormalTok{ r\_forest}\SpecialCharTok{$}\FunctionTok{do\_predict}\NormalTok{(X)}
\NormalTok{  sigma }\OtherTok{\textless{}{-}}\NormalTok{ r\_forest}\SpecialCharTok{$}\FunctionTok{get\_sigma}\NormalTok{()}
  
  \ControlFlowTok{for}\NormalTok{(i }\ControlFlowTok{in} \DecValTok{1}\SpecialCharTok{:}\NormalTok{num\_iter) \{}
\NormalTok{    R }\OtherTok{\textless{}{-}}\NormalTok{ y }\SpecialCharTok{{-}}\NormalTok{ r}
\NormalTok{    beta }\OtherTok{\textless{}{-}} \FunctionTok{update\_beta}\NormalTok{(R, Z, sigma}\SpecialCharTok{\^{}}\DecValTok{2}\NormalTok{)}
\NormalTok{    eta }\OtherTok{\textless{}{-}} \FunctionTok{as.numeric}\NormalTok{(X }\SpecialCharTok{\%*\%}\NormalTok{ beta)}
\NormalTok{    R }\OtherTok{\textless{}{-}}\NormalTok{ y }\SpecialCharTok{{-}}\NormalTok{ eta}
\NormalTok{    r }\OtherTok{\textless{}{-}}\NormalTok{ r\_forest}\SpecialCharTok{$}\FunctionTok{do\_gibbs}\NormalTok{(X, R, X, }\DecValTok{1}\NormalTok{)}
\NormalTok{    sigma }\OtherTok{\textless{}{-}}\NormalTok{ r\_forest}\SpecialCharTok{$}\FunctionTok{get\_sigma}\NormalTok{()}
    
\NormalTok{    r\_out[i,] }\OtherTok{\textless{}{-}}\NormalTok{ r}
\NormalTok{    beta\_out[i,] }\OtherTok{\textless{}{-}}\NormalTok{ beta\_out}
\NormalTok{    sigma\_out[i] }\OtherTok{\textless{}{-}}\NormalTok{ sigma}
\NormalTok{    eta\_out[i,] }\OtherTok{\textless{}{-}}\NormalTok{ eta}
\NormalTok{  \}}
  
  \FunctionTok{return}\NormalTok{(}\FunctionTok{list}\NormalTok{(}\AttributeTok{r =}\NormalTok{ r\_out, }\AttributeTok{beta =}\NormalTok{ beta\_out, }\AttributeTok{sigma =}\NormalTok{ sigma\_out, }
              \AttributeTok{eta =}\NormalTok{ eta\_out, }\AttributeTok{mu =}\NormalTok{ eta\_out }\SpecialCharTok{+}\NormalTok{ r\_out))}
    
\NormalTok{\}}
\end{Highlighting}
\end{Shaded}

  \bibliography{mybib.bib}

\end{document}